\documentclass[fleqn,usenatbib]{mnras}

\usepackage{newtxtext,newtxmath}

\usepackage[T1]{fontenc}
\usepackage{float}

\DeclareRobustCommand{\VAN}[3]{#2}
\let\VANthebibliography\thebibliography
\def\thebibliography{\DeclareRobustCommand{\VAN}[3]{##3}\VANthebibliography}

\usepackage{graphicx}	
\usepackage{amsmath}	
\usepackage{multirow}

\usepackage{soul}
\usepackage{orcidlink}
\usepackage{hyperref}

\newcommand{\ndet}{462}
\newcommand{\nstars}{9,821}
\newcommand{\nmatching}{290}
\newcommand{\nlos}{2,714}

\newcommand{\kms}{km s$^{-1}$}

\def\13co{$^{13}$CO}
\def\aj{AJ}

\def\aap{A\&A}
\def\apj{ApJ}
\def\apjs{ApJS}
\def\pasa{PASA}

\def\c18o{C$^{18}$O}

\def\cm2{cm$^{-2}$}
\def\cm3{cm$^{-3}$}

\def\FCNM{$f_\mathrm{CNM}$}

\def\h2{H$_2$}
\def\hi{H{\sc i}}
\def\ki{K{\sc i}}

\def\CaII{Ca~{\sc ii}}

\def\NaI{Na~{\sc i}}
\def\kms{$\mathrm{km\,s^{-1}}$}
\def\k0{$\kappa_{0}$}

\def\nat{Nature}
\def\n2h{N$_2$H$^+$}
\def\NHI{$N_\mathrm{H{\sc I}}$}
\def\NHICNM{$N_\mathrm{H{\sc I},CNM}$}
\def\NHIWNM{$N_\mathrm{H{\sc I},WNM}$}

\def\nhiUnit{$\times\ 10^{20}$ cm$^{-2}$}
\def\NHIthin{$N^{*}_\mathrm{HI}$}

\def\NH{$N_\mathrm{H}$}

\def\NKI{$N_\mathrm{K{\sc I}}$}
\def\EWKI{$EW_\mathrm{K{\sc I}}$}

\def\s{s$^{-1}$}
\def\s353{$\sigma_{353}$}

\def\taupeak{$\tau_{\mathrm{peak}}$}

\def\TBpeak{$T_{\mathrm{b,peak}}$}

\def\Ts{$T_{\mathrm{s}}$}
\def\t353{$\tau_{353}$}

\title[GASKAP \& GALAH: Interstellar \hi\ and \ki\ absorption towards the foreground of the Magellanic Clouds]{Multi-wavelength probes of the Milky Way's Cold Interstellar Medium: Radio \hi\ and Optical \ki\ Absorption with GASKAP and GALAH}

\author[H. Nguyen et al.]{Hiep Nguyen\orcidlink{0000-0002-2712-4156}$^{1}$\thanks{E-mail: \href{mailto:vanhiep.nguyen@anu.edu.au}{vanhiep.nguyen@anu.edu.au}},
Sven Buder\orcidlink{0000-0002-4031-8553}$^{1,2}$, Juan D. Soler\orcidlink{0000-0002-0294-4465}$^{3}$, N.~M. McClure-Griffiths\orcidlink{0000-0003-2730-957X}$^{1}$,
J. R.  Dawson\orcidlink{0000-0003-0235-3347}$^{4,5}$, \newauthor
James Dempsey\orcidlink{0000-0002-4899-4169}$^{1,6}$,
Helga Dénes\orcidlink{0000-0002-9214-8613}$^{7}$, 
John M. Dickey\orcidlink{0000-0002-6300-7459}$^{8}$,
Ian Kemp\orcidlink{0000-0002-6637-9987}$^{9}$,
Denis Leahy\orcidlink{0000-0002-4814-958X}$^{10}$,
Min-Young Lee\orcidlink{0000-0002-9888-0784}$^{11}$, \newauthor
Callum Lynn\orcidlink{0000-0001-6846-5347}$^{1}$,
Yik Ki Ma\orcidlink{0000-0003-0742-2006}$^{12}$,
Antoine Marchal\orcidlink{0000-0002-5501-232X}$^{1}$,
Marc-Antoine Miville-Desch\^{e}nes\orcidlink{0000-0002-7351-6062}$^{13}$, \newauthor
Eric G. M. Muller\orcidlink{0000-0001-5621-1577}$^{1}$, Claire E. Murray\orcidlink{0000-0002-7743-8129}$^{14,15}$, Gyueun Park\orcidlink{0000-0002-7374-7864}$^{11}$,
Nickolas M. Pingel\orcidlink{0000-0001-9504-7386}$^{16}$,
Hilay Shah\orcidlink{0000-0002-9136-6731}$^{1}$, \newauthor
Sne\v{z}ana Stanimirovi\'{c}\orcidlink{0000-0002-3418-7817}$^{16}$, Jacco Th. van Loon\orcidlink{0000-0002-1272-3017}$^{17}$
\\
$^{1}$Research School of Astronomy and Astrophysics, The Australian National University, Canberra, ACT 2611, Australia\\
$^{2}$ARC Centre of Excellence for All Sky Astrophysics in 3 Dimensions (ASTRO 3D), Australia\\
$^{3}$Istituto di Astrofisica e Planetologia Spaziali (IAPS). INAF. Via Fosso del Cavaliere 100, 00133 Roma, Italy\\
$^{4}$Department of Physics and Astronomy and MQ Research Centre in Astronomy, Astrophysics, and Astrophotonics, Macquarie University, NSW 2109, Australia\\
$^{5}$Australia Telescope National Facility, CSIRO Space and Astronomy, PO Box 76, Epping NSW 1710, Australia\\
$^{6}$CSIRO Information Management and Technology, GPO Box 1700 Canberra, ACT 2601, Australia\\
$^{7}$School of Physical Sciences and Nanotechnology, Yachay Tech University, Hacienda San Jos\'e S/N, 100119, Urcuqu\'i, Ecuador\\
$^{8}$School of Natural Sciences, Private Bag 37, University of Tasmania, Hobart, TAS, 7001, Australia\\
$^{9}$International Centre for Radio Astronomy Research (ICRAR), Curtin University, Bentley, WA 6102, Australia\\
$^{10}$Department of Physics and Astronomy, University of Calgary, Calgary, AB, Canada T2N 1N4\\
$^{11}$Korea Astronomy and Space Science Institute, 776 Daedeok-daero, Daejeon 34055, Republic of Korea\\
$^{12}$Max-Planck-Institut f\"ur Radioastronomie, Auf dem H\"ugel 69, 53121 Bonn, Germany\\
$^{13}$AIM, CEA, CNRS, Université Paris-Saclay, Universit\'{e} Paris Diderot, Sorbonne Paris Cit\'{e}, F-91191 Gif-sur-Yvette, France\\
$^{14}$Department of Physics \& Astronomy, Johns Hopkins University, 3400 N. Charles Street, Baltimore, MD 21218, USA\\
$^{15}$Space Telescope Science Institute, 3700 San Martin Drive, Baltimore, MD 21218, USA\\
$^{16}$Department of Astronomy, University of Wisconsin-Madison, 475 N Charter St, Madison, WI 53703, USA\\
$^{17}$Lennard-Jones Laboratories, Keele University, ST5 5BG, UK
}

\date{Accepted XXX. Received YYY; in original form ZZZ}

\pubyear{2025}

\begin{document}
\label{firstpage}
\pagerange{\pageref{firstpage}--\pageref{lastpage}}
\maketitle

\begin{abstract}
We present a comparative
analysis of interstellar hydrogen (\hi) and potassium (\ki) absorption from the radio and optical surveys, GASKAP and GALAH, to study the physical and kinematic properties of the cold interstellar medium (ISM) in the Milky Way foreground towards the Magellanic Clouds. By comparing GASKAP \hi\ absorption with interstellar \ki\ absorption detected in GALAH spectra of nearby stars (within 12 arcmin angular distance or a spatial separation of $\sim$0.75 pc), we reveal a strong kinematic correlation between these two tracers of the cold neutral ISM. The velocity offsets between matched \hi\ and \ki\ absorption components are small, with a mean (median) offset of $-$1.3 ($-$1.2) \kms\ and a standard deviation of 2.3 \kms. The high degree of kinematic consistency suggests a close spatial association between \ki\ and cold \hi\ gas. Correlation analyses reveal a moderate positive relationship 
between \hi\ and \ki\ line-of-sight properties, such as \ki\ column density with \hi\ column density or \hi\ brightness temperature. We observe a $\sim$63\% overlap  in the detection of both species towards \nmatching\ (out of 462) GASKAP \hi\ absorption lines of sight, and estimate a median \ki/\hi\ abundance ratio of $\sim$2.3$\times$10$^{-10}$, in excellent agreement with previous findings. Our work opens up an exciting avenue of Galactic research that uses large-scale surveys in the radio and optical wavelengths to probe the neutral interstellar medium through its diverse tracers.
\end{abstract}

\begin{keywords}
ISM: atoms -- ISM: general -- ISM: abundances -- ISM: lines and bands -- radio lines: ISM
\end{keywords}



\section{Introduction}
    Understanding the structure, composition, and dynamics of the neutral interstellar medium (ISM) is central to unraveling the processes that govern star formation and galaxy evolution \citep[e.g.][]{Draine2011,Kennicutt2012}. The neutral ISM is predominantly composed of atomic hydrogen (\hi) and various trace elements, including metals such as potassium (\ki), which serve as important diagnostics of its multiphase structure. \hi\ gas, observable via its 21-cm hyperfine transition, is the primary tracer of neutral atomic material, while neutral potassium, detected through optical absorption lines near 7699 \r{A}, selectively probes the denser and colder regions of the atomic ISM \citep[][]{Trapero1995,Welty2001}. Together, these tracers can provide a more detailed characterisation of the physical conditions, chemical enrichment, and small-scale structure of the atomic ISM.

The neutral atomic hydrogen in the ISM exists in two main thermally stable forms: the warm neutral medium (WNM) with temperature $\sim$8000 K and density $\sim$0.5 \cm3, and the cold neutral medium (CNM) with temperature $\sim$100 K and density $\sim$10 \cm3, for typical Solar metallicity \citep{Field1969,McKee1977,Wolfire2003}. While \hi\ emission generally traces the bulk of the atomic gas across a wide range of temperatures and densities, \hi\ absorption preferentially traces the CNM at lower temperatures and higher densities \citep[][]{Heiles2003a,Lee2015,Murray2018,Kalberla2018,Marchal2019,Nguyen2019,McClure-Griffiths2023,Nguyen2024}.

Neutral potassium \ki\ provides an additional probe of the cold neutral medium. With a low ionisation potential \citep[4.34 eV,][]{Corliss1979}, \ki\ atoms are more readily photoionised than sodium (\NaI) or calcium (\CaII), resulting in a lower abundance -- typically by a factor of $\sim$15 compared to \NaI\ and \CaII\ \citep{Welty2001}. Consequently, significantly higher total hydrogen column densities \NH\ are required for \ki\ to be detectable, limiting its presence to dense, well-shielded regions where the CNM typically resides \citep{Hobbs1974,Welty2001}. This sensitivity to high column density and shielding makes \ki\ an effective tracer of the cold, dense components of the atomic ISM. High-resolution observations of the \ki\ 7699 \r{A} resonance line are particularly useful for resolving interstellar component structures in cases where the more abundant \NaI\ lines become saturated. For instance, interstellar \ki\ absorption has been used to investigate cool \hi\ gas within the local ISM  \citep{Trapero1995}, and to explore small-scale spatial structure in high column density environments \citep{Lauroesch1999}.

Recent large-scale surveys have revolutionised our ability to study ISM components in detail. In this work, we examine the Galactic ISM from the synergy between two major Galactic surveys: the Galactic Australian Square Kilometre Array Pathfinder survey \citep[GASKAP;][]{Dickey2013,Pingel2022} and the Galactic Archaeology with HERMES survey \citep[GALAH;][]{DeSilva2015,Buder2024}. The GASKAP-\hi\ survey in radio provides unprecedented sensitivity and angular resolution (30 arcsec) in mapping both \hi\ emission and absorption in the direction of the Magellanic System, including Large and Small Magellanic Clouds, Magellanic Bridge, and Magellanic Stream \citep[e.g.,][]{McClureGriffiths2018NatAs,DiTeodoro2019,Szotkowski2019,Dempsey2020,Dempsey2022,Ma2023,Gerrard2023,Murray2024,Chen2025,BucklandWillis2025} as well as their Milky Way foreground \citep[e.g.,][]{Dickey2022,Nguyen2024,Lynn2025}. The GALAH optical spectroscopic survey, meanwhile, offers extensive measurements of interstellar \ki\ absorption features toward a large sample of background stars. The combination of these data sets presents a unique opportunity to study the spatial and kinematic relations between total \hi, cold \hi\ (CNM traced by \hi\ absorption) and the dense neutral ISM traced by interstellar \ki\ absorption. Namely, we will investigate (\textit{i}) the kinematic relationship between \hi\ and \ki\ absorption, and (\textit{ii}) the abundance of neutral potassium relative to neutral hydrogen.

We focus on the \hi\ gas at high Galactic latitudes in the Milky Way foreground towards the Magellanic Clouds (MCs) ($b$ $\sim$ $-$45$^{\circ}$ to $-$25$^{\circ}$). This region is characterised by distinct filamentary structures, composed of gas and dust, as illustrated in Figure \ref{fig:all_src_locations}. These filamentary structures are likely to reside close to the surface of the Local Bubble, a low-density gas-filled region formed by stellar feedback \citep{Berkhuijsen1971, Cox1987, McKee1998, Zucker2022}, at distances $\sim$200 to 300 pc from the Sun \citep{ONeill2024,Erceg2024}. In this MC foreground, GASKAP-\hi's large field of view ($5 \times 5$ square degrees) facilitates simultaneous measurements of emission and absorption towards \nlos\ background continuum sources across a 250 square degree area \citep{Nguyen2024}.

In this work, we utilise \hi\ properties derived from GASKAP-\hi\ absorption measurements with the Pilot II Magellanic Cloud \hi\ foreground observations \citep[or the GASKAP local \hi\ survey,][]{Nguyen2024} and \ki\ properties obtained from GALAH Data Release 4 \citep[GALAH DR4,][]{Buder2024}. Section \ref{sec:observations} outlines the observational data sets from GASKAP and GALAH surveys. We then examine the kinematic relationship between interstellar \ki\ and \hi\ absorption in Section \ref{sec:ki_hi_kinematics}. In Section \ref{sec:integrated_props}, we discuss \hi\ and \ki\ integrated properties and the \ki\ abundances in the local ISM. Finally, Section \ref{sec:conclusions} summarises our findings.

\section{Observational data}
    \label{sec:observations}

\subsection{GASKAP-\hi\ absorption and emission in radio spectra}
\label{subsec:hi_abs_em}

\begin{figure*}
 \center
  \includegraphics[width=\textwidth]{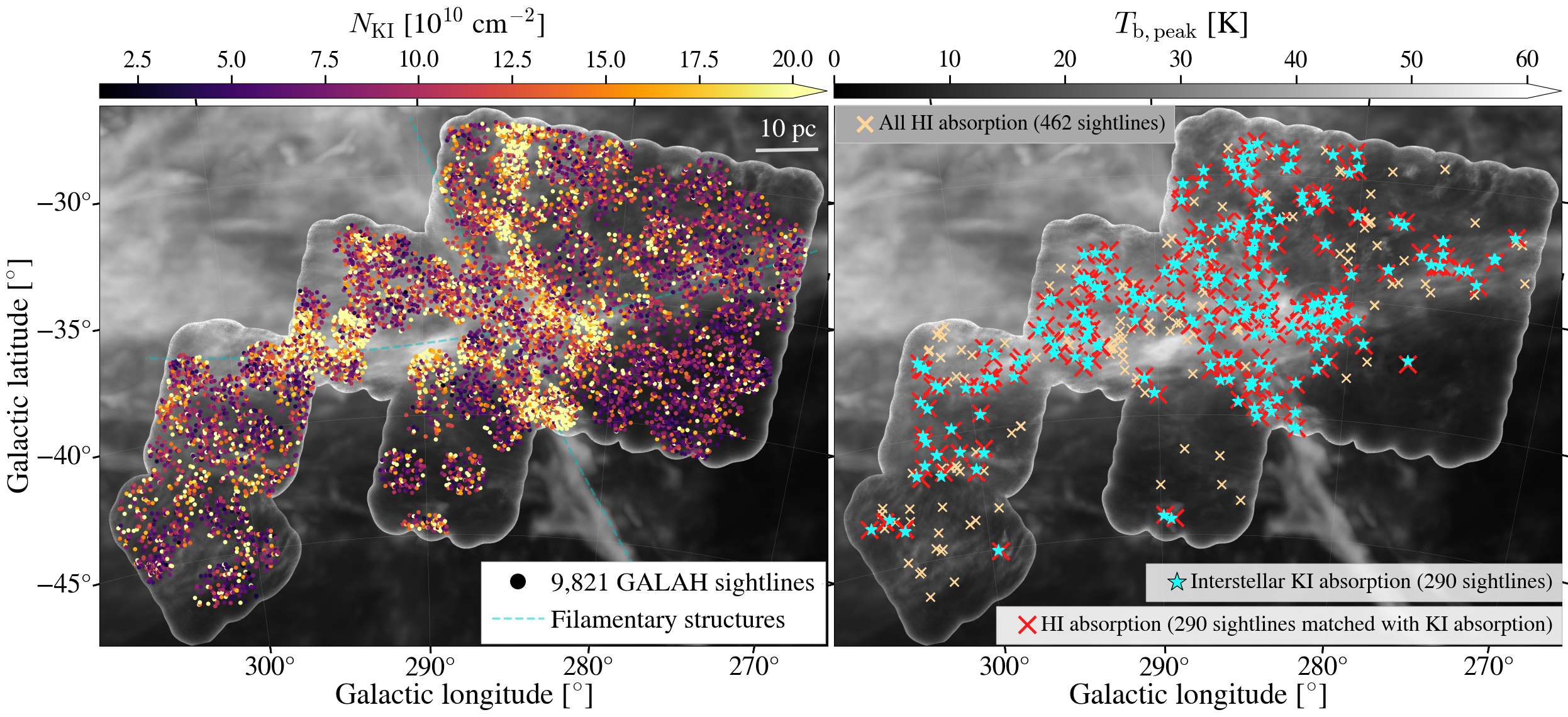}
  \caption{\textit{Left panel}: On-sky spatial distribution of GALAH \ki\ column density measurements towards \nstars\ selected stars in the Magellanic Cloud foreground, overlaid on the GASKAP peak brightness temperature map (\TBpeak, with an angular resolution of 30 arcsec). The gray outer boundary marks the GASKAP-\hi\ 250-square-degree observing footprint, whereas the background outside this footprint shows the Parkes Murriyang single-dish peak brightness from the Galactic All-Sky Survey \citep[][]{McClureGriffiths2009}. Two local \hi\ filaments are depicted as gray dashed curves. \textit{Right panel}: Locations of \ndet\ background radio continuum sources ($S_\mathrm{1.4\ GHz} \geq$ 15 mJy) in the GASKAP-\hi\ Pilot II survey (orange crosses) with \hi\ detections \citep{Nguyen2024}, overlaid on the GASKAP \hi\ peak brightness temperature map \TBpeak\ (identical to the background displayed in the left panel). A subset of \nmatching\ \hi\ absorption sightlines matched with \ki\ absorption detections is highlighted with red crosses. Corresponding \nmatching\ GALAH stars with interstellar \ki\ detections are shown as star markers. An interactive 3D view is available at \href{https://nv-hiep.github.io/gaga/}{github.io/gaga}.}\vspace{0.4cm}
  \label{fig:all_src_locations}
\end{figure*}

We utilise \hi\ absorption and emission data from the GASKAP local \hi\ survey's on-/off-source measurements \citep{Nguyen2024}. This survey covered an area of $\sim$250 square degrees in the direction of Magellanic Clouds, spanning a Galactic latitude range of $b$ = ($-45^{\circ}$, $-25^{\circ}$) and a Galactic longitude range of $l$ = ($270^{\circ}$, $305^{\circ}$). The GASKAP-\hi\ absorption pipeline developed by \cite{Dempsey2022} was employed to extract \hi\ absorption spectra for \nlos\ sources. To enhance signal-to-noise sensitivity while still resolving cold \hi\ gas structures, the Milky Way GASKAP absorption data extending over a velocity range from $-50$ to $+50$ km s$^{-1}$, initially acquired with a native spectral resolution of 0.244 km s$^{-1}$, were smoothed to a spectral resolution of 0.977 km s$^{-1}$. Local \hi\ absorption features were detected at the 3$\sigma$ threshold towards \ndet\ continuum background sources. \cite{Nguyen2024} then performed a Gaussian decomposition on the paired emission-absorption spectra to derive key \hi\ physical properties: peak optical depths (\taupeak), central velocities ($v_\mathrm{0}$), velocity dispersions ($\Delta v_\mathrm{FWHM}$) of individual absorbing gas components, spin temperatures (\Ts) of the gas, total \hi\ column densities (\NHI), CNM column densities (\NHICNM), WNM column densities (\NHIWNM), and CNM fractions (\FCNM). Please refer to \cite{Nguyen2024} for further details.

All interstellar \ki\ and \hi\ emission and absorption spectra, along with their kinematic properties, \hi\ and \ki\ column densities, their associated uncertainties, and the data analysis notebooks used in this study, are publicly available at \href{https://doi.org/10.5281/zenodo.17211553}{DOI 10.5281/17211553}.

\begin{table*}
\begin{center}
\fontsize{8}{7}\selectfont
\caption{Basic information of \nmatching\ lines of sight with \hi\ and \ki\ detections (full table for the sources is available in the online version). For each GASKAP source, we list its Galactic longitudes/latitudes ($l$, $b$) as well as the nearby stellar Galactic longitudes/latitudes ($l^{\star}$, $b^{\star}$), the star's distance to the Sun ($D^{\star}$), \hi\ and \ki\ column densities (\NHI, \NKI), CNM central velocities ($V_\mathrm{CNM}$), the corresponding matched \ki\ absorption velocities ($V_\mathrm{KI}$), and \hi-\ki\ velocity offsets ($\Delta V_\mathrm{HI-KI}$).}
\centering
\label{table:source_list}
\begin{tabular}{llllllcccccc}
\noalign{\smallskip} \hline \hline \noalign{\smallskip}
\shortstack{Source\\ $\ $} & 
\shortstack{$l$\\$(^{o})$} & \shortstack{$b$\\$(^{o})$} & \shortstack{$l^{\star}$\\$(^{o})$} & \shortstack{$b^{\star}$\\$(^{o})$} & \shortstack{$D^{\star}$\\(pc)} & \shortstack{\NHI \\($10^{20}$ cm$^{-2}$)} & \shortstack{\NKI \\($10^{10}$ cm$^{-2}$)} & \shortstack{$V_\mathrm{CNM}$ \\(\kms)} & \shortstack{$V_\mathrm{KI}$ \\(\kms)} & \shortstack{$\Delta V_\mathrm{HI-KI}$ \\(\kms)}\\
\hline

J002337$-$735529 & 305.56 & $-$43.06 & 305.33 & $-$43.02 & 1170.7 & 2.9 & 3.9 & $-$0.57 & 5.04 & $-$5.61 \\
J003414$-$733327 & 304.61 & $-$43.51 & 304.60 & $-$43.60 & 1482.7 & 3.0 & 6.2 & 2.88 & 3.29 & $-$0.41 \\
J004741$-$753010 & 303.25 & $-$41.62 & 303.45 & $-$41.50 & 214.5 & 3.7 & 27.3 & $-$0.99 & 1.52 & $-$2.51 \\
J005019$-$755629 & 303.02 & $-$41.19 & 302.92 & $-$41.14 & 1176.1 & 4.5 & 18.8 & 9.27 & 8.26 & 1.01 \\
J005321$-$770019 & 302.79 & $-$40.12 & 302.65 & $-$40.17 & 632.5 & 4.4 & 5.6 & 0.09 & $-$6.85 & 6.94 \\
J005341$-$771713 & 302.77 & $-$39.84 & 302.77 & $-$39.89 & 593.3 & 4.2 & 6.7 & 0.13 & 2.16 & $-$2.03 \\
J005641$-$783945 & 302.60 & $-$38.46 & 302.62 & $-$38.54 & 1689.8 & 5.2 & 16.2 & $-$2.12 & $-$1.85 & $-$0.27 \\
J010120$-$781900 & 302.29 & $-$38.80 & 302.30 & $-$38.84 & 1300.2 & 5.1 & 9.2 & $-$2.26 & $-$4.24 & 1.98 \\
J010214$-$801239 & 302.36 & $-$36.91 & 302.43 & $-$36.91 & 608.4 & 4.9 & 8.9 & 3.01 & 4.19 & $-$1.18 \\
J010249$-$795604 & 302.31 & $-$37.18 & 302.31 & $-$37.10 & 850.9 & 5.2 & 8.6 & 3.76 & 4.65 & $-$0.89 \\
J010251$-$753523 & 301.98 & $-$41.52 & 302.16 & $-$41.65 & 3235.1 & 4.7 & 33.0 & 9.89 & 9.39 & 0.50 \\
J010452$-$795246 & 302.19 & $-$37.23 & 302.09 & $-$37.21 & 759.9 & 6.0 & 8.1 & 3.64 & 4.07 & $-$0.43 \\
J010912$-$790840 & 301.87 & $-$37.95 & 301.70 & $-$37.94 & 442.0 & 5.9 & 10.9 & 2.15 & 6.98 & $-$4.83 \\
J011549$-$771147 & 301.17 & $-$39.85 & 301.23 & $-$39.80 & 1926.0 & 4.6 & 10.7 & $-$0.04 & $-$0.53 & 0.49 \\
J011552$-$761226 & 301.01 & $-$40.83 & 300.96 & $-$40.81 & 1723.3 & 4.2 & 8.0 & 7.33 & 8.38 & $-$1.05 \\
J011623$-$790554 & 301.44 & $-$37.96 & 301.51 & $-$38.08 & 667.8 & 5.5 & 10.2 & 3.52 & 6.10 & $-$2.58 \\
J012250$-$715042 & 299.48 & $-$45.07 & 299.60 & $-$45.04 & 1880.1 & 4.0 & 14.9 & $-$1.04 & 2.09 & $-$3.13 \\
J012257$-$751507 & 300.25 & $-$41.71 & 300.21 & $-$41.60 & 340.2 & 4.3 & 55.2 & $-$10.25 & $-$8.30 & $-$1.95 \\
J013252$-$760405 & 299.65 & $-$40.80 & 299.69 & $-$40.96 & 670.3 & 3.6 & 9.6 & 1.28 & 2.05 & $-$0.77 \\
J013910$-$784115 & 299.98 & $-$38.17 & 300.18 & $-$38.21 & 3520.5 & 5.8 & 6.9 & 3.54 & 3.29 & 0.25 \\
J013926$-$772856 & 299.59 & $-$39.33 & 299.63 & $-$39.38 & 1516.3 & 4.4 & 13.7 & 5.09 & 3.54 & 1.55 \\
J014601$-$783125 & 299.51 & $-$38.24 & 299.56 & $-$38.29 & 2007.4 & 6.3 & 8.1 & 2.06 & 3.61 & $-$1.55 \\
J015552$-$782819 & 298.89 & $-$38.15 & 298.78 & $-$38.02 & 608.0 & 5.6 & 10.9 & 2.80 & $-$0.58 & 3.38 \\
J015714$-$784232 & 298.90 & $-$37.90 & 299.02 & $-$37.88 & 808.0 & 6.4 & 8.9 & 2.27 & 2.21 & 0.06 \\
J020448$-$795503 & 298.99 & $-$36.64 & 299.00 & $-$36.66 & 2603.3 & 5.8 & 31.7 & 5.23 & 5.12 & 0.11 \\
J020803$-$792116 & 298.57 & $-$37.12 & 298.47 & $-$36.99 & 2033.4 & 6.0 & 12.7 & 4.00 & 4.32 & $-$0.32 \\
J021929$-$781445 & 297.38 & $-$37.92 & 297.60 & $-$37.98 & 928.6 & 6.1 & 16.3 & $-$1.74 & $-$1.90 & 0.16 \\
J022458$-$784143 & 297.31 & $-$37.39 & 297.33 & $-$37.45 & 465.8 & 8.0 & 23.9 & 2.53 & 4.30 & $-$1.77 \\
J022934$-$784745 & 297.12 & $-$37.20 & 297.06 & $-$37.24 & 787.3 & 7.0 & 16.4 & 2.16 & 3.14 & $-$0.98 \\
J024508$-$695753 & 289.80 & $-$44.15 & 289.86 & $-$44.18 & 879.1 & 3.8 & 16.4 & 8.58 & 12.59 & $-$4.01 \\
J024800$-$693600 & 289.21 & $-$44.29 & 289.43 & $-$44.30 & 636.1 & 3.3 & 16.2 & 9.35 & 9.88 & $-$0.53 \\
J025415$-$791927 & 296.21 & $-$36.15 & 296.15 & $-$36.19 & 1056.7 & 6.2 & 20.5 & 2.24 & 2.57 & $-$0.33 \\
J025618$-$780044 & 295.17 & $-$37.17 & 295.39 & $-$37.18 & 650.9 & 7.0 & 11.6 & 1.98 & 2.80 & $-$0.82 \\
J025721$-$782714 & 295.43 & $-$36.77 & 295.34 & $-$36.84 & 858.3 & 7.3 & 29.5 & 2.65 & 4.50 & $-$1.85 \\
J025803$-$783711 & 295.52 & $-$36.62 & 295.65 & $-$36.57 & 2661.4 & 6.5 & 16.2 & 1.97 & 3.02 & $-$1.05 \\
J030316$-$765256 & 293.94 & $-$37.85 & 293.73 & $-$37.75 & 418.7 & 5.8 & 7.8 & 1.85 & $-$0.42 & 2.27 \\
J030320$-$791455 & 295.73 & $-$35.96 & 295.89 & $-$35.89 & 698.8 & 6.7 & 16.6 & 1.93 & 2.24 & $-$0.31 \\
J030330$-$772932 & 294.39 & $-$37.36 & 294.33 & $-$37.38 & 938.5 & 7.4 & 19.9 & $-$0.05 & 0.92 & $-$0.97 \\
J031155$-$765150 & 293.44 & $-$37.55 & 293.66 & $-$37.54 & 1760.2 & 6.2 & 16.6 & 3.92 & 4.83 & $-$0.91 \\
J031235$-$782909 & 294.71 & $-$36.29 & 294.62 & $-$36.25 & 583.5 & 6.0 & 29.9 & 1.71 & 4.42 & $-$2.71 \\
J031836$-$795958 & 295.64 & $-$34.95 & 295.51 & $-$35.02 & 761.8 & 5.9 & 7.5 & 1.92 & 1.93 & $-$0.01 \\
J032019$-$773939 & 293.66 & $-$36.65 & 293.77 & $-$36.66 & 1021.6 & 7.0 & 25.0 & 2.57 & 4.59 & $-$2.02 \\
J032259$-$770023 & 292.97 & $-$37.03 & 293.08 & $-$36.90 & 441.7 & 7.7 & 39.4 & 5.77 & 4.68 & 1.09 \\
J032301$-$770029 & 292.97 & $-$37.03 & 293.16 & $-$37.01 & 1049.8 & 8.0 & 32.6 & 2.58 & 3.41 & $-$0.83 \\
J032456$-$775225 & 293.62 & $-$36.33 & 293.61 & $-$36.42 & 707.5 & 6.7 & 12.7 & 2.38 & 5.06 & $-$2.68 \\
J032550$-$735329 & 290.04 & $-$39.13 & 290.23 & $-$39.05 & 1351.3 & 5.5 & 31.1 & 8.35 & 5.59 & 2.76 \\
J032704$-$794916 & 295.16 & $-$34.83 & 295.34 & $-$34.91 & 216.7 & 5.9 & 9.7 & 2.37 & 6.14 & $-$3.77 \\
J032753$-$774732 & 293.41 & $-$36.28 & 293.48 & $-$36.21 & 637.6 & 8.1 & 30.3 & 2.44 & 3.83 & $-$1.39 \\
J032834$-$745053 & 290.75 & $-$38.33 & 290.73 & $-$38.43 & 445.9 & 9.9 & 21.7 & 4.78 & 7.18 & $-$2.40 \\
J033037$-$784908 & 294.17 & $-$35.44 & 294.19 & $-$35.57 & 1677.0 & 6.1 & 21.9 & 3.57 & 6.13 & $-$2.56 \\

\noalign{\smallskip} \hline \noalign{\smallskip}


\end{tabular}
\end{center}
\end{table*}

\subsection{GALAH DR4: Interstellar \ki\ absorption in optical spectra}
\label{subsec:galah_ki_absorption}

The Galactic Archaeology with HERMES (GALAH) Survey Data Release 4 \citep{Buder2024} provides measurements of high-resolution optical stellar spectra from the High Efficiency and Resolution Multi-Element Spectrograph (HERMES) via the 2dF facility at the 3.9-m Anglo-Australian Telescope \citep{Lewis2002,Miszalski2006,Barden2010,Brzeski2011,Heijmans2012,Farrell2014,Sheinis2015}. HERMES operates with a nominal resolving power of $R$ = 28,000 and covers four bands of the optical wavelength range (471$-$490, 565$-$587, 648$-$674, and 759$-$789 nm). The spectral coverage includes several key ISM absorption features: the atomic \ki\ absorption line at 7698.9643\r{A} as well as diffuse interstellar band features at 5780.59\r{A}, 5797.19\r{A}, and 6613.66\r{A} \citep{Vogrincic2023}. The wavelength calibration is performed by fitting a polynomial function to the identified peak positions of the Thorium-Xenon (ThXe) emission lines, obtained from dedicated arc-lamp calibration exposures \citep{Kos2017}. For the fourth band that includes the \ki\ line, up to 31 peaks can be detected with typical root mean square values of the wavelength solution around 0.027 \r{A}. At the \ki\ wavelength of 7699 \r{A}, the spectral resolution corresponds to 0.27 \r{A} (or $\sim$10.7 \kms\ in velocity space). The observations are over-sampled on the CCD with a spacing of only 0.0736 \r{A} (equivalent to 2.87 \kms). The wavelength calibration of optical spectra -- without the use of more sophisticated methods like laser frequency combs -- is typically much less precise than those of radio observations.

The analysis pipeline of GALAH DR4 fits synthetic stellar spectra, interpolated via neural networks, to the whole wavelength range in order to extract stellar parameters and elemental abundances of up to 32 elements for 917,588 stars, while iteratively shifting the observed spectrum with a global stellar radial velocity.

In modeling stellar spectra, the line spread function (accounting for instrumental and grating effects) and the rotational velocity of a star (which determines stellar rotational broadening) are taken into account. For slow rotators -- particularly giant stars, which generally exhibit low rotational velocities -- the intrinsic stellar \ki\ absorption line remains narrow. As a result, it can closely resemble interstellar \ki\ absorption. For cases where the stellar and ISM line-of-sight velocities overlap, the fitting can become degenerate and lead to overestimated stellar \ki\ abundances while no ISM \ki\ can be detected in the residuals. To address this issue and disentangle stellar absorption features from interstellar counterparts, the GALAH DR4 analysis by \cite{Buder2024} implements a post-processing step in which Gaussian profiles are fitted to the residuals between observed and synthetic spectra in selected spectral regions, including those containing interstellar absorption components. The fit provides not only the central wavelength to recover the line-of-sight velocity, but also the amplitude and line width (Full Width at Half Maximum, FWHM) which can be converted into an equivalent width ($EW_\mathrm{KI}$). Please refer to \citet{Buder2024} for further details. We estimate the noise level in the \ki\ absorption spectra as the standard deviation ($\sigma_\mathrm{KI}$) of off-line wavelength regions.

We then select stars inside the GASKAP-\hi\ observing footprint with (1) high \ki\ signal-to-noise: residual flux amplitudes ($F_\mathrm{KI,resid}$, the difference between the observed data and stellar model) exceeding $3\sigma_\mathrm{KI}$ (three times the noise level) and \ki\ equivalent width $EW_\mathrm{KI} >$ 0.004 \r{A}; (2) ages older than 100 Myr to exclude (circum-)stellar contributions to interstellar \ki\ content from very young stars; and (3) stellar distances $D_{\star} >$ 300 pc, corresponding to the approximate distance to the Local Bubble shell \citep[e.g.,][and also see Appendix \ref{app_distances}]{ONeill2024,Leike2020}. These selection criteria result in a sample of \nstars\ stars across the region of interest towards the Magellanic Cloud foreground (as shown in the left panel of Figure \ref{fig:all_src_locations}). This stellar set serves as the basis for the subsequent analysis presented in this paper.

\section{Kinematic relationship between interstellar \ki\ and \hi\ absorption}
    \label{sec:ki_hi_kinematics}

We examine the kinematic relationship between CNM and interstellar \ki\ absorption features by combining data from the GASKAP-\hi\ survey (462 \hi\ absorption-detected lines of sight) and the GALAH DR4 catalogue (\nstars\ selected stars, as detailed in Section \ref{subsec:galah_ki_absorption}). It is worth noting that \cite{Buder2024} fitted \ki\ absorption with a single Gaussian, while \cite{Nguyen2024} modelled the \hi\ absorption spectra with multiple CNM Gaussian components. To establish reliable matches and enable meaningful comparisons between these two datasets, we consider both spatial (on-sky) and spectral properties.

To enable direct spectral comparison, we converted the GALAH stellar spectra from wavelength to Doppler velocity in the kinematic Local Standard of Rest (LSRK) frame, using the Solar peculiar motion $(U_\odot, V_\odot, W_\odot)$ = (11.10, 12.24, 7.25) \kms\ \citep[][]{Schonrich2010},  where $U_\odot$ is directed towards the Galactic center, $V_\odot$ is in the direction of Galactic rotation, and $W_\odot$ points towards the north Galactic pole. This transformation ensures consistency with the velocity frame adopted in the GASKAP-\hi\ observations. For spatial matching in the plane of sky, we adopted a search radius of 12 arcmin around each GASKAP-\hi\ absorption sightline to identify candidate stars from the GALAH DR4 survey. Assuming a typical distance of 220 pc (see Appendix \ref{app_distances}) to the absorbing gas, this angular radius corresponds to a distance separation of $\sim$0.75 pc. This value represents a balance between maintaining physical association with the absorbing gas and retaining a statistically significant number of matched stars. Along the GASKAP-\hi\ sightlines, the number of nearby GALAH stars varies between three and 14. In addition, we extended the search radius to 24 arcmin around each GASKAP-\hi\ absorption sightline (corresponding to a linear scale of 1.5 pc) to examine how the \ki-\hi\ relation between \ki\ and \hi\ varies with spatial matching. In this case, the number of nearby GALAH stars per GASKAP-\hi\ absorption detection ranges from three to 50.

Among the stars located within the 0.75 pc search radius around each GASKAP \hi\ absorption line of sight, we select only the stars with clear separation between stellar and interstellar \ki\ absorption features $\Delta V^\mathrm{\star}_\mathrm{KI} >$ 11 \kms\ (approximately equal to the GALAH \ki\ spectral resolution). If multiple stars meet the criteria, the best-matched star is chosen as the one that simultaneously minimises the velocity offset between the interstellar \ki\ absorption feature and the nearest \hi\ absorption Gaussian component $|\Delta V_\mathrm{CNM-KI}|$ and maximises the \ki\ absorption amplitude. In cases where two nearby GASKAP-\hi\ absorption sightlines are matched to the same star, we treat each sightline–star association independently, while noting that the interstellar \ki\ absorption feature is the same in both cases. Applying this procedure, we identify \nmatching\ out of 462 GASKAP-\hi\ lines of sight with confident detections of both \hi\ and interstellar \ki\ absorption, yielding a matching interstellar absorption detection rate of $\sim$63\%. When the matching radius is increased to 1.5 pc around each GASKAP-\hi\ absorption sightline, the number of detections of both \hi\ and interstellar \ki\ absorption rises to 403 out of 462 sightlines, yielding a detection rate of $\sim$87\%. We will focus primarily on the 290-sightline sample with smaller search radius of 0.75 pc, while using the 403-sightline sample for comparison.

We show in Figure \ref{fig:all_src_locations}'s right panel the positions of \nmatching\ matching lines of sight with both \hi\ and interstellar \ki\ absorption detections (red crosses), overlaid on the GASKAP-\hi\ peak brightness temperature map in Galactic coordinates. An interactive 3D view is available at \href{https://nv-hiep.github.io/gaga/}{github.io/gaga}. Table \ref{table:source_list} summarises their basic information, including Galactic longitudes/latitudes ($l$, $b$), corresponding stars' Galactic longitudes/latitudes ($l^{\star}$, $b^{\star}$), distance to star ($D^{\star}$), \hi\ and \ki\ column densities (\NHI, \NKI), CNM central velocities ($V_\mathrm{CNM}$), the corresponding matched \ki\ absorption velocities ($V_\mathrm{KI}$), and \hi-\ki\ velocity offsets ($\Delta V_\mathrm{HI-KI}$) for a sample of 50 lines of sight. A full table containing information for \nmatching\ matching \hi-\ki\ lines of sight is available in the online version of this publication.

\begin{figure}
\includegraphics[width=1.0\linewidth]{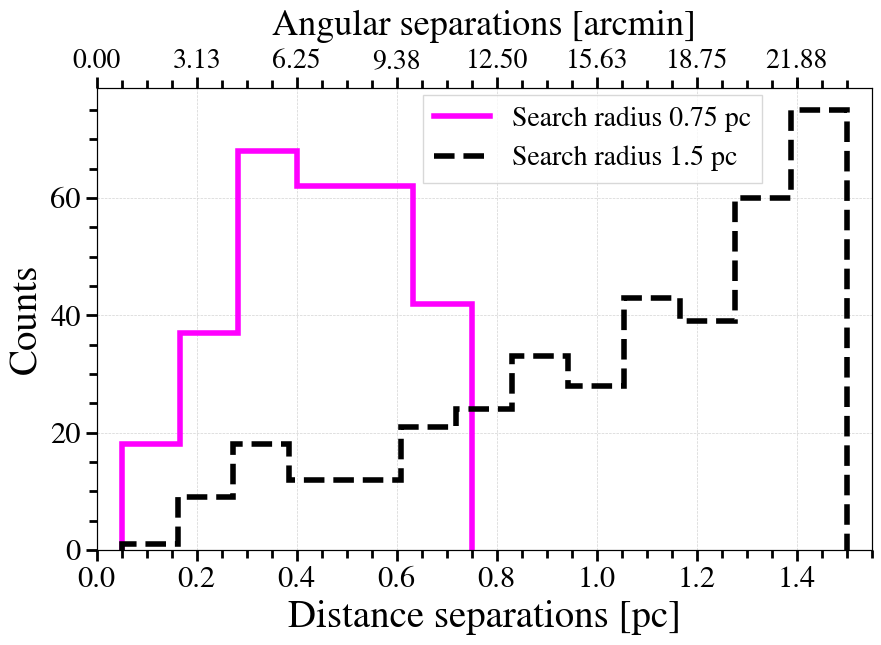}
\caption{Distributions of angular and distance separations between the GASKAP-\hi\ absorption sightlines and their corresponding matched stars. The solid line represents a sample of 290 lines of sight selected by a search radius of 0.75 pc around a GASKAP-\hi\ absorption detection, while the dashed line represents the sample of 403 lines of sight selected within a 1.5 pc radius.}
\label{fig:distance_seps}
\end{figure}

Figure \ref{fig:distance_seps} shows the distributions of the angular and distance separations between the GASKAP-\hi\ absorption sightlines and their corresponding matched stars. The solid line represents a sample of 290 sightlines with both \hi\ and \ki\ absorption identified within a 0.75 pc search radius around a GASKAP-\hi\ absorption detection, while the dashed line denotes a sample of 403 sightlines obtained using a 1.5 pc radius. For the 290-sightline sample, the angular separations range from 0.45 to 11.7 arcmin. Assuming a distance of 220 pc to the local absorbing \hi\ gas (as discussed above), these translate to physical separations of 0.03 -- 0.75 pc, with a mean, median and standard deviation of 0.44 pc, 0.44 pc and 0.17 pc, respectively.

\begin{figure}
\includegraphics[width=1.0\linewidth]{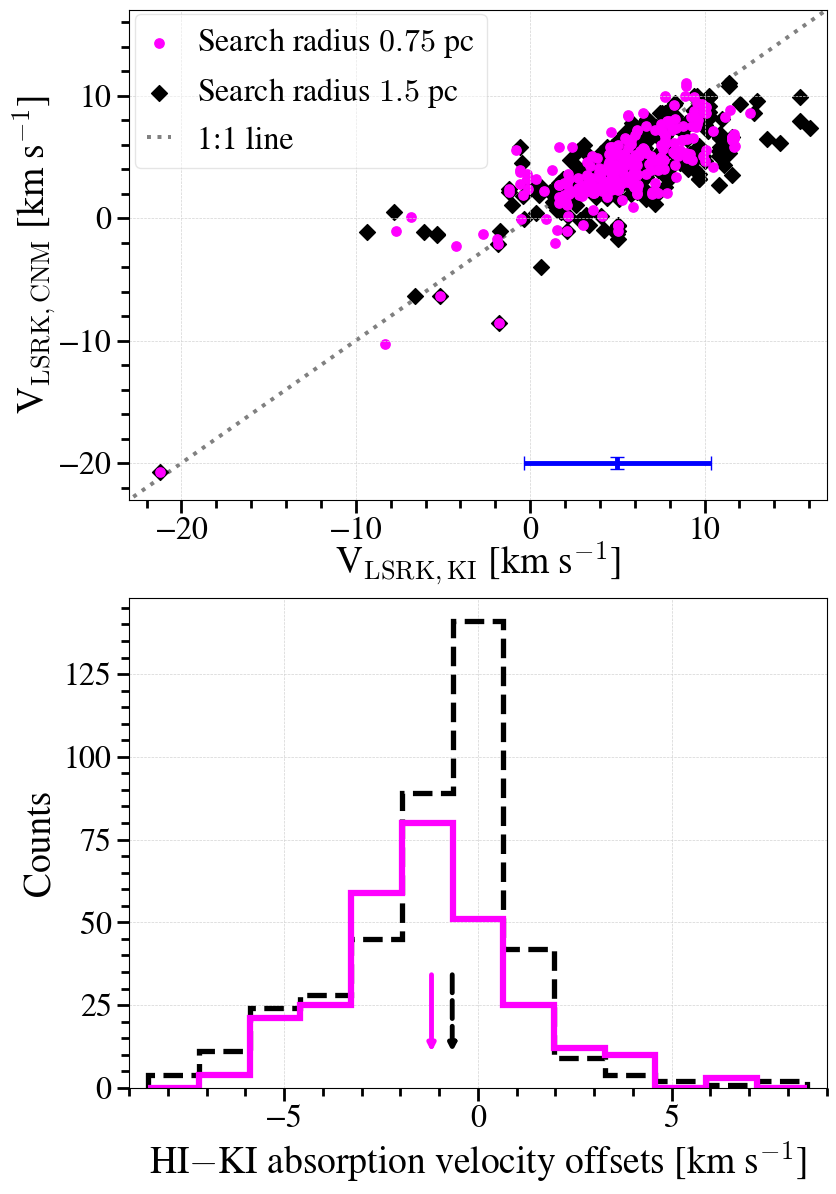}
\caption{\textit{Top panel}: Comparison of CNM and \ki\ absorption velocities for 290 sightlines (magenta), where both \hi\ and \ki\ absorption were identified within a 0.75 pc search radius around a GASKAP-\hi\ absorption detection, and for 403 sightlines (black) using a 1.5 pc radius. Error bars represent the velocity uncertainties: 0.5 \kms\ for CNM and 5.35 \kms\ (half the spectral resolution) for \ki. \textit{Bottom panel}: Histogram of velocity offsets between CNM and \ki\ absorption features ($V_\mathrm{CNM} - V_\mathrm{KI}$): solid line for the 290-sightline sample and dashed line for the 403-sightline sample. Arrows mark their median offsets, $-1.2$ \kms\ and $-0.7$ \kms, respectively.}
\label{fig:hi_ki_vlsrk}
\end{figure}

\begin{figure}
\includegraphics[width=1.0\linewidth]{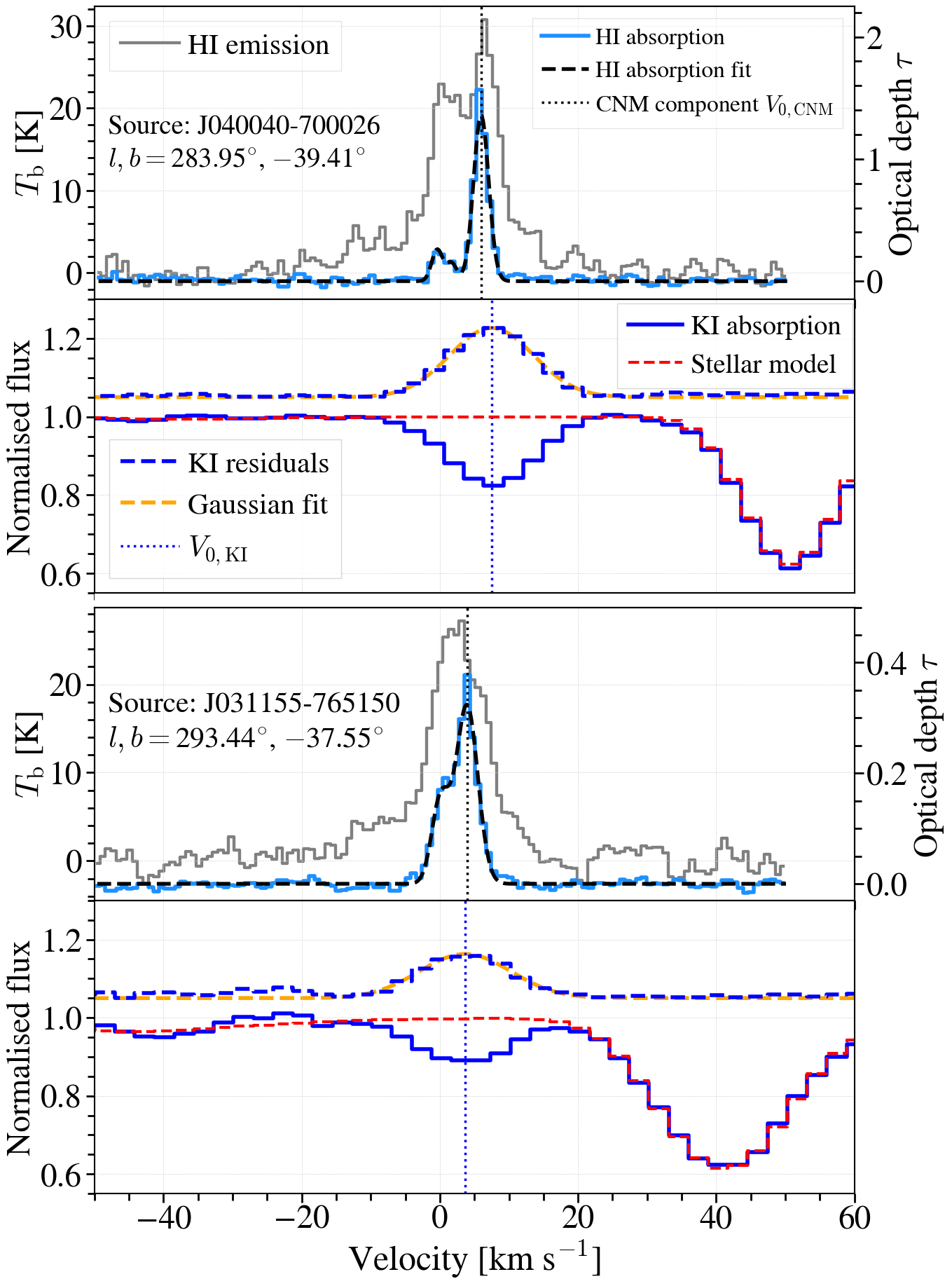}
\caption{\hi/\ki\ sample spectra of the Magellanic Cloud foreground towards two background continuum sources J040040--700026 and J031155--765150. \textit{Top panels}: \hi\ emission (grey), absorption (light blue), and a Gaussian fit to the \hi\ absorption profile (black dashed line). The central velocity of a CNM component is indicated by the vertical dotted line. \textit{Bottom panels}: GALAH stellar spectrum (blue solid line), synthetic stellar spectrum (red dashed line), residual \ki\ absorption (blue dashed line), and a Gaussian fit to the \ki\ residual (orange dashed line). The vertical dotted line indicates the central velocity of the \ki\ absorption component, which aligns with the CNM central velocity. More samples are available at \href{https://doi.org/10.5281/zenodo.17211553}{DOI 10.5281/17211553}.}
\label{fig:ki_hi_spectra}
\end{figure}

The kinematic correlation between the central velocities of the \ki\ and \hi\ absorption components is shown in Figure \ref{fig:hi_ki_vlsrk}'s top panel, with their velocity offsets $\Delta V_\mathrm{CNM-KI}$ (here $V_\mathrm{0, CNM} - V_\mathrm{0, KI}$) displayed in the lower panel. The uncertainties in CNM velocities, derived from the Gaussian decomposition of GASKAP-\hi\ spectra, are typically lower than 0.25 \kms\ \citep{Nguyen2024}; but for simplicity, here we assume a uniform maximum uncertainty of 0.5 \kms\ (half of the \hi\ channel width). For the \ki\ velocities, we adopt the GALAH \ki\ spectral resolution of 10.7 \kms\ as a reference, leading to an estimated uncertainty of $\sim$5.35 \kms\ per individual measurement (i.e., half the resolution). These velocity uncertainties are illustrated as error bars in the upper panel of Figure \ref{fig:hi_ki_vlsrk}.

A strong correlation between \hi\ and \ki\ velocities in our sample, with Pearson (Spearman) coefficients at 0.79 (0.73) and $p-$values $\sim$ 10$^{-62}$ ($\sim$ 10$^{-49}$) implies a close kinematic association between interstellar \ki\ and CNM (see examples in Figure \ref{fig:ki_hi_spectra}), indicating their co-movement and co-existence in the Galactic ISM. These results are broadly consistent with \cite{Liszt2021}, who compared higher-resolution \ki\ spectra (R $\sim$ 100,000 or $\sim$3 \kms) obtained with the UV-Visual Echelle Spectrograph on the Very large Telescope, with \hi\ absorption \citep{Dickey1983} observed with the Very Large Array along five compact extragalactic continuum sources. Using a kinematic correlation coefficient defined as the velocity-space overlap integral of the absorption profiles, \cite{Liszt2021} found values around 0.7 (see their Table 4), also indicative of a strong kinematic correspondence. Although the correlation metrics differ, both studies reveal that \hi\ and \ki\ absorption features are closely aligned in velocity space.

As shown in the lower panel of Figure \ref{fig:hi_ki_vlsrk}, the \hi-\ki\ velocity offset distribution for the 290-sightline sample (solid line) spans from $-$7 \kms\ to $+$7 \kms, with a mean of $-$1.3 \kms, a median of $-$1.2 \kms, and a standard deviation of 2.3 \kms. For the 403-sightline sample (dashed line), the \hi–\ki\ velocity offsets span $-$8.5 to $+$8.3, with a mean of $-$1.1 \kms, a median of $-$0.7 \kms, and a standard deviation of 2.3 \kms. Kolmogorov-Smirnov tests for both samples indicate that the $\Delta V_\mathrm{CNM-KI}$ distribution is inconsistent with a Gaussian centered at zero (KS = 0.31, $p-$value $\sim$ 10$^{-24}$ and KS = 0.34, $p-$value $\sim$ 10$^{-42}$, respectively), suggesting a small kinematic difference between the CNM and \ki. However, given the relatively large uncertainties in the \ki\ central velocities arising from its original spectral resolution $\sim$10.7 \kms, we believe that there is no evidence to support real physical differences. While such differences may exist, the current \ki\ spectral resolution does not allow us to comment on them at this level. Since the \ki\ lines of sight are identified as close to those of \hi\ absorption, and if we interpret velocity as a proxy for spatial location (i.e., CNM components at different velocities correspond to different distances), then our results implies that \ki\ is likely located within or near CNM environments.

We note that whenever \hi\ absorption is undetected along the 2,239 local GASKAP-\hi\ sightlines \citep{Nguyen2024}, \ki\ absorption features also tend to be absent above the 3$\sigma$ threshold. Additionally, after degrading the \hi\ spectral resolution to match that of the \ki\ spectra, we find that interstellar \ki\ absorption features are consistently narrower than the \hi\ emission lines, but generally comparable in width to the \hi\ absorption features. A higher-resolution spectroscopic study by \cite{Liszt2021}, using \ki\ absorption spectra with $\sim$3 \kms\ resolution towards five \hi\ absorption sightlines, found that \ki\ absorption features are slightly narrower than the corresponding \hi\ absorption features (see their Figure 1). In that work, the author analysed the line profiles of five species (\hi, HCO$^+$, \ki, Na \textsc{i}, and Ca \textsc{ii}) along the same lines of sight, and concluded that HCO$^+$ exhibits the narrowest lines, followed in order by \ki, \hi, Na \textsc{i}, and Ca \textsc{ii}.

\section{Integrated properties: \ki\ vs \hi}
    \label{sec:integrated_props}

In this Section we explore quantitative relationships between \ki\ and \hi\ column densities as well as the \ki/\hi\ abundance. We define the relative abundance of neutral potassium \ki\ with respect to neutral hydrogen \hi\ as the ratio of their respective integrated column densities, \NKI/\NHI. \hi\ column density is obtained from the GASKAP absorption survey, and therefore is corrected for opacity effects \citep[see][]{Nguyen2024}. The \ki\ column density is derived from the \ki\ equivalent width (listed in GALAH DR4 as \texttt{ew\_k\_is}), which is measured by Gaussian fits to the residuals between the GALAH observed spectra and the corresponding stellar models (as depicted by the blue dashed lines in Figure \ref{fig:ki_hi_spectra}). Subsequently, assuming the interstellar \ki\ line to be optically thin, we can employ the linear regime of the curve of growth, which establishes a direct proportionality between the \ki\ column density \NKI\ and its equivalent width \EWKI:
\begin{equation}
    N_\mathrm{KI} \frac{\pi e^2 f \lambda^2}{m_e c^2} = \int (-\ln r_\nu)d\nu \approx EW_{\mathrm{KI}},
    \label{eq:ki_ew}
\end{equation}
with \ki\ wavelength $\lambda$ at 7698.9643\r{A}, residual intensity $r_\nu$ for a frequency $\nu$, as well as electron charge $e$, electron mass $m_e$, speed of light $c$, and oscillator strength $f$ \citep{Hobbs1974b}. For the \ki\ the resonance transition between the first excited electronic state to the ground electronic state (3p$^6$4p $^2$P$_\mathrm{3/2}$ $\rightarrow$ 3p$^6$4s $^2$S$_\mathrm{1/2}$), we adopted a logarithm of the oscillator strength $\log gf = -0.178$ from \citet{Trubko2017}. Given the statistical weight of the lower level ($g = 2J + 1 = 2$ for the 3p$^6$4s $^2$S$_\mathrm{1/2}$ state with a total angular momentum $J = 1/2$) and the provided $\log gf$ value, this corresponds to an oscillator strength of $f = 0.332$.

We note that while \citet[][]{Buder2024} provided values for the \ki\ equivalent width ($EW_{\text{KI}}$), \ki\ flux amplitude ($A_{\text{KI}}$), and the standard deviation ($\sigma_{\text{KI}}$) from Gaussian fits to the \ki\ absorption features, they did not report the corresponding uncertainties for these quantities. To estimate the uncertainties of the equivalent widths and the resulting column densities, we assume that the relative uncertainties for both \ki\ flux amplitude and the standard deviation are 10\%. We then propagate these assumed uncertainties to estimate the uncertainty of the equivalent width, $\delta EW_{\text{KI}}$, using the standard error propagation formula. This results in an estimated uncertainty of $\sim$14\% for the equivalent widths. We then use this uncertainty to estimate the uncertainty in the derived \ki\ column densities.

Below, we perform correlation analyses using both the \citet[][]{Hobbs1974} dataset (a small sample of ten lines of sight with relatively broad Galactic longitude coverage) and our own dataset consisting of \nmatching\ matched lines of sight towards the Magellanic Cloud foreground. This comparison allows us to assess the consistency and statistical significance of any potential trends between \ki\ and \hi\ column densities. In addition, we extend our analysis to a larger sample of \nstars\ selected stars within the GASKAP-\hi\ field of view by comparing GALAH \ki\ column densities with GASKAP \hi\ peak brightness temperatures and column densities under the optically-thin assumption.

\begin{figure}
\centering
\includegraphics[width=1.0\linewidth]{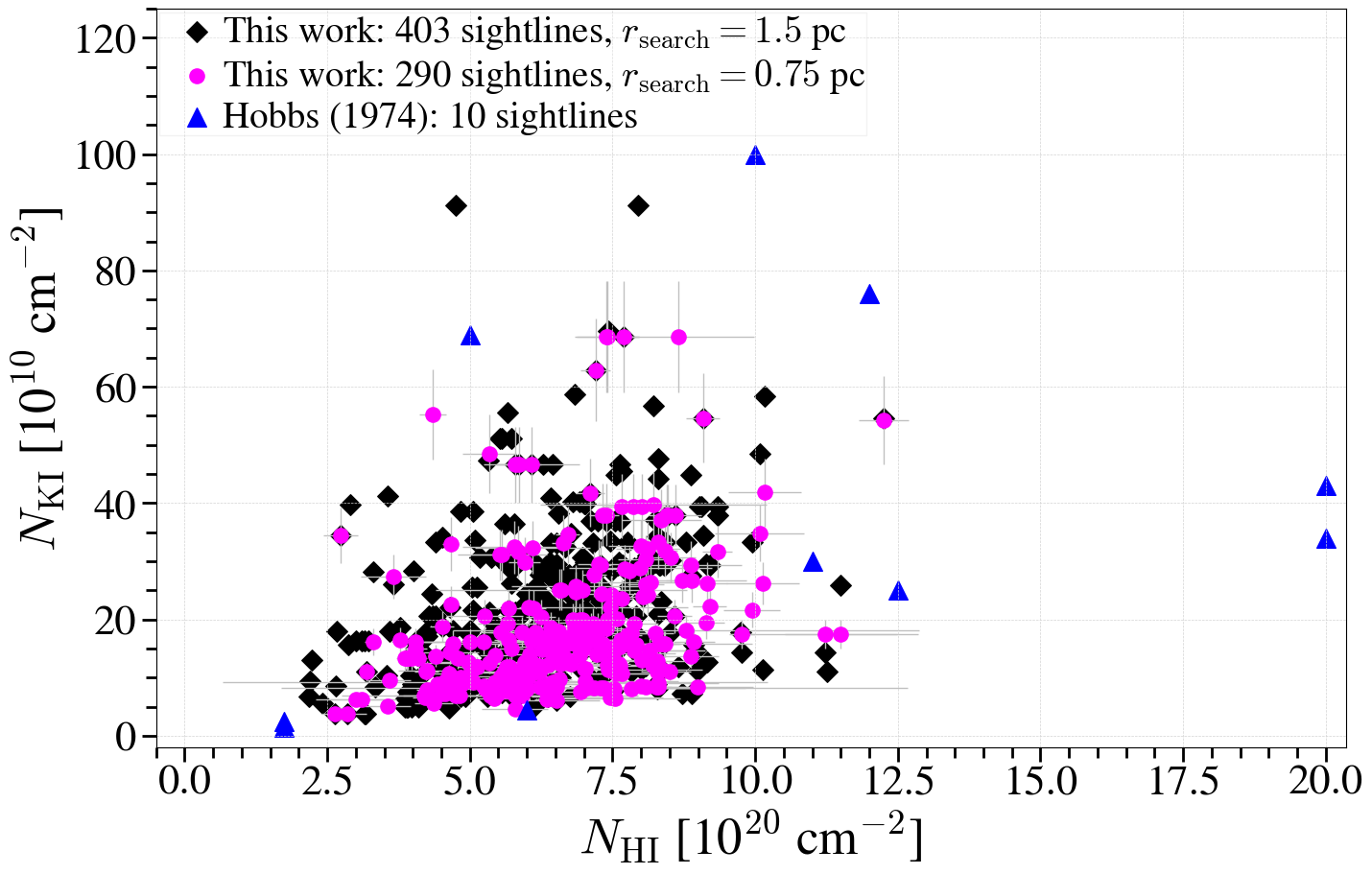}
\caption{Opacity-corrected \hi\ vs \ki\ column densities. Circles represent our sample for \nmatching\ sightlines with a search radius of $r_\mathrm{search} =$ 0.75 pc, diamonds show the 403-sightline sample with a 1.5 pc search radius, and triangles indicate \citet{Hobbs1974}'s sample of ten lines of sight.}
\label{fig:corr_hi_ki_relation}
\end{figure}



\subsection{\ki-\hi\ relation}
\label{subsec:ki_abundance}

We first examine the \ki\ and \hi\ column density values reported by \citet[][triangle markers in Figure \ref{fig:corr_hi_ki_relation}]{Hobbs1974}. \citet[][]{Hobbs1974} determined their opacity-corrected \hi\ column densities from Lyman$-\alpha$ absorption measurements obtained with the Orbiting Astronomical Observatory 2 \citep[OAO-2,][]{Savage1972}. Since Lyman$-\alpha$ absorption traces the total number of \hi\ atoms, it provides the ``true'' \hi\ column density along a line of sight, independent of spin temperature. Pearson and Spearman correlation analyses for their ten sightlines yield coefficients of 0.31 ($p-$value = 0.38) and 0.45 ($p-$value = 0.20), respectively. While both coefficients show positive trends, their associated $p-$values are well above the conventional significance threshold of 0.05. Due to their small sample size, we cannot confidently conclude the presence of a statistically significant linear or monotonic relationship between \ki\ and \hi\ based on \citet{Hobbs1974} data, as the observed trends could readily be attributed to random chance within such a limited dataset.

In contrast, our correlation analyses along \nmatching\ matched sightlines (within a 0.75 pc search radius) reveal a moderate positive relationship between opacity-corrected \hi\ and \ki\ column densities (indicated by circle markers in Figure \ref{fig:corr_hi_ki_relation}). The Pearson correlation coefficient is $r = 0.36$, whereas the Spearman coefficient is $\rho = 0.43$. The correlation tests are highly statistically significant, with small $p-$values of 5 $\times$ 10$^{-10}$ and 4 $\times$ 10$^{-14}$, respectively. Using a larger sample of 403 sightlines (within a 1.5 pc search radius), we also find a moderate positive correlation between \NKI\ and \NHI, though with slightly lower coefficients ($r = 0.34, \rho = 0.40$) and corresponding $p-$values of 10$^{-12}$ and 10$^{-17}$. The statistical significance in both cases, supported by substantial sample sizes (\nmatching\ and 403 sightlines), suggests a robust, though modest positive association between \hi\ and \ki. For comparison, \cite{Liszt2021} found a Pearson correlation coefficient of 0.31 (see their Table 3) between \NKI\ and \NHI\ along five sightlines with both absorption detections, broadly consistent with our results.

We also search for the relations between \NKI\ and other absorption-based tracers of cold \hi\ gas. We found a moderate positive correlation between \NKI\ and the CNM optical depth (\(\tau_{\mathrm{CNM}}\) of the CNM components associated with the \ki\ absorption feature). 
The Pearson correlation coefficient is $r = 0.29$ with a highly significant $p-$value $=$ \(8 \times 10^{-7}\), while the Spearman rank coefficient is 
\(\rho = 0.22\) with $p-$value $=$ \(1.5 \times 10^{-4}\). A similarly strong relationship is observed between \NKI\
and the line-of-sight CNM column density (\(N_{\mathrm{HI,\,CNM}}\)). 
Here, the Pearson coefficient is \(r = 0.27\) with 
$p-$value $=$ \(3 \times 10^{-6}\), and the Spearman coefficient is 
\(\rho = 0.25\) with $p-$value $=$ \(1.6 \times 10^{-5}\). The relationship between \NKI\ and the \hi\ integrated optical depth (\(\Upsilon_{\mathrm{HI}}\)) is slightly stronger with \(r = 0.26\) ($p-$value $=$ \(1.4 \times 10^{-5}\)) and \(\rho = 0.27\) ($p-$value $=$ \(2.2 \times 10^{-7}\)).

\begin{figure}
\centering
\includegraphics[width=1.0\linewidth]{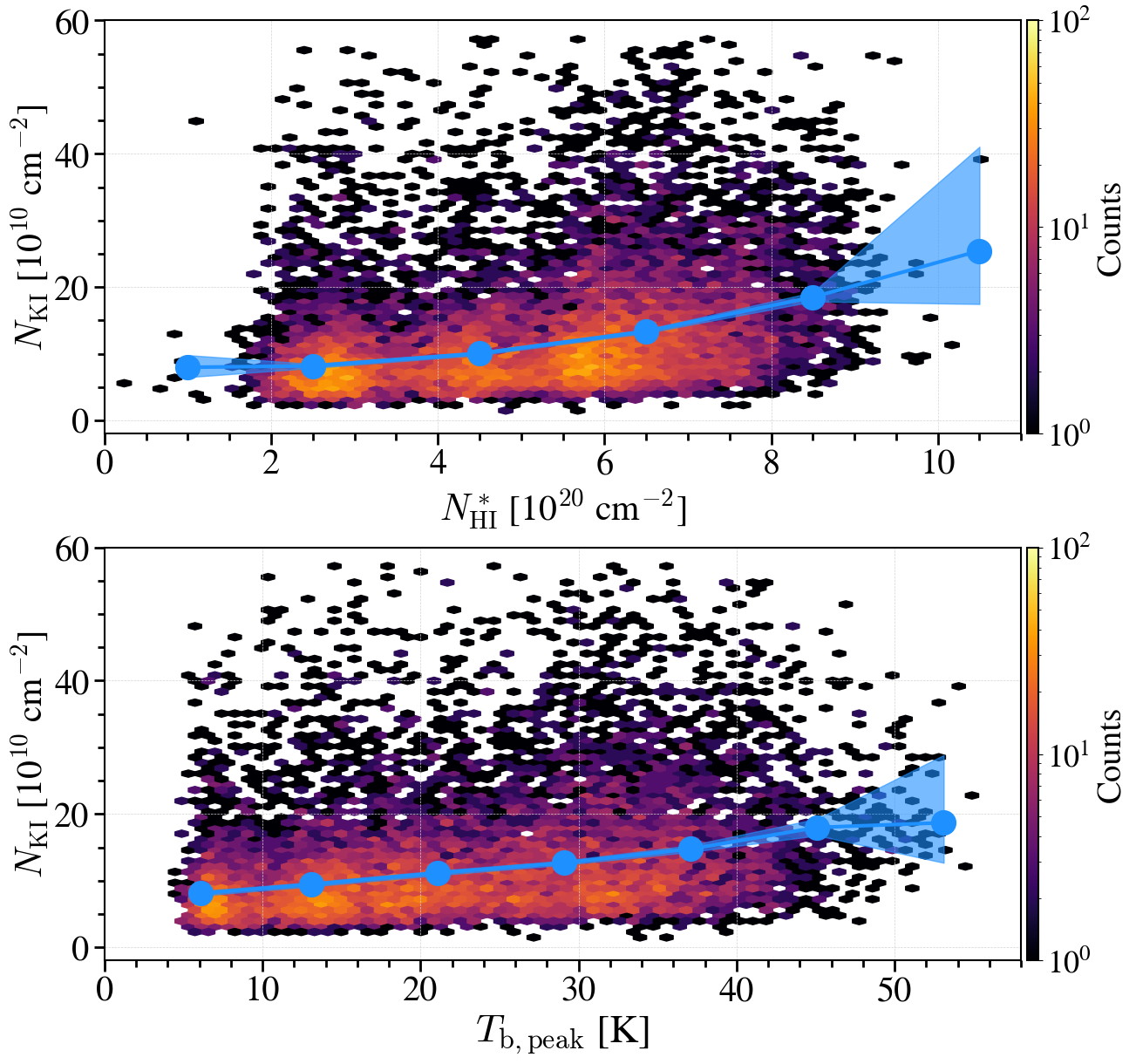}
\caption{Correlations between \hi\ and \ki\ line-of-sight properties towards \nstars\ stars in the Magellanic Cloud foreground. \textit{Top panel}: \ki\ column density (\NKI) versus optically-thin \hi\ column density (\NHIthin). \textit{Bottom panel}: \ki\ column density versus \hi\ peak brightness temperature (\TBpeak). Circles and shaded regions denote median values and their 1$\sigma$ uncertainties per \NHIthin\ and \TBpeak\ bins, respectively. Color indicates data point density.}
\label{fig:hi_ki_relation}
\end{figure}

We then further examine the \hi-\ki\ line-of-sight properties for a larger sample of \nstars\ selected sightlines (within the GASKAP-\hi\ observing footprint) using optically-thin \hi\ column density \NHIthin, peak brightness temperature \TBpeak\ from \hi\ emission data, and \ki\ column density \NKI. An on-sky distribution of \ki\ column densities towards \nstars\ stars, as shown in the left panel of Figure \ref{fig:all_src_locations}, reveals a distinctive spatial behavior. While the \ki\ column density distribution nicely follows the morphology of \hi\ filamentary structures characterised by high column densities (\NHIthin\ $>$ 5 \nhiUnit) and peak brightness temperature (\TBpeak\ $>$ 30 K), it also scatters across the map. Figure \ref{fig:hi_ki_relation} compares \NKI\ with both \NHIthin\ and \TBpeak; here, circle markers and shaded regions represent median values and 1$\sigma$ uncertainties in column density and brightness temperature bins.

In these subsequent broader analyses, both Pearson and Spearman correlation tests indicate positive relationships between \hi\ and \ki\ properties. For the \NHIthin-\NKI\ pair, the Pearson correlation coefficient is 0.30 ($p-$value $\ll 10^{-50}$), 
while the Spearman rank correlation is slightly higher at 0.37 ($p-$value $\ll 10^{-50}$). 
The correlation between \NKI\ and \TBpeak\ is similarly positive, with a Pearson coefficient of 0.28 ($p-$value $\ll 10^{-50}$), 
and a Spearman coefficient of 0.33 ($p-$value $\ll 10^{-50}$). 
The high statistical significance (indicated by extremely low $p-$values, effectively zero) across the latter analyses confirms that these positive \hi-\ki\ correlations are unlikely to be due to random chance. In all \ki-\hi\ column density comparisons carried out in this Section, the Spearman correlation coefficients are higher than their respective Pearson coefficients. This suggests that while \NKI\ tends to increase as \NHI\ (or \TBpeak) increases, their relationships might be better described by a general increasing trend rather than a linear one.

\subsection{\ki/\hi\ abundance}
\label{subsec:ki_abundance}

\begin{figure}
\centering
\includegraphics[width=1.0\linewidth]{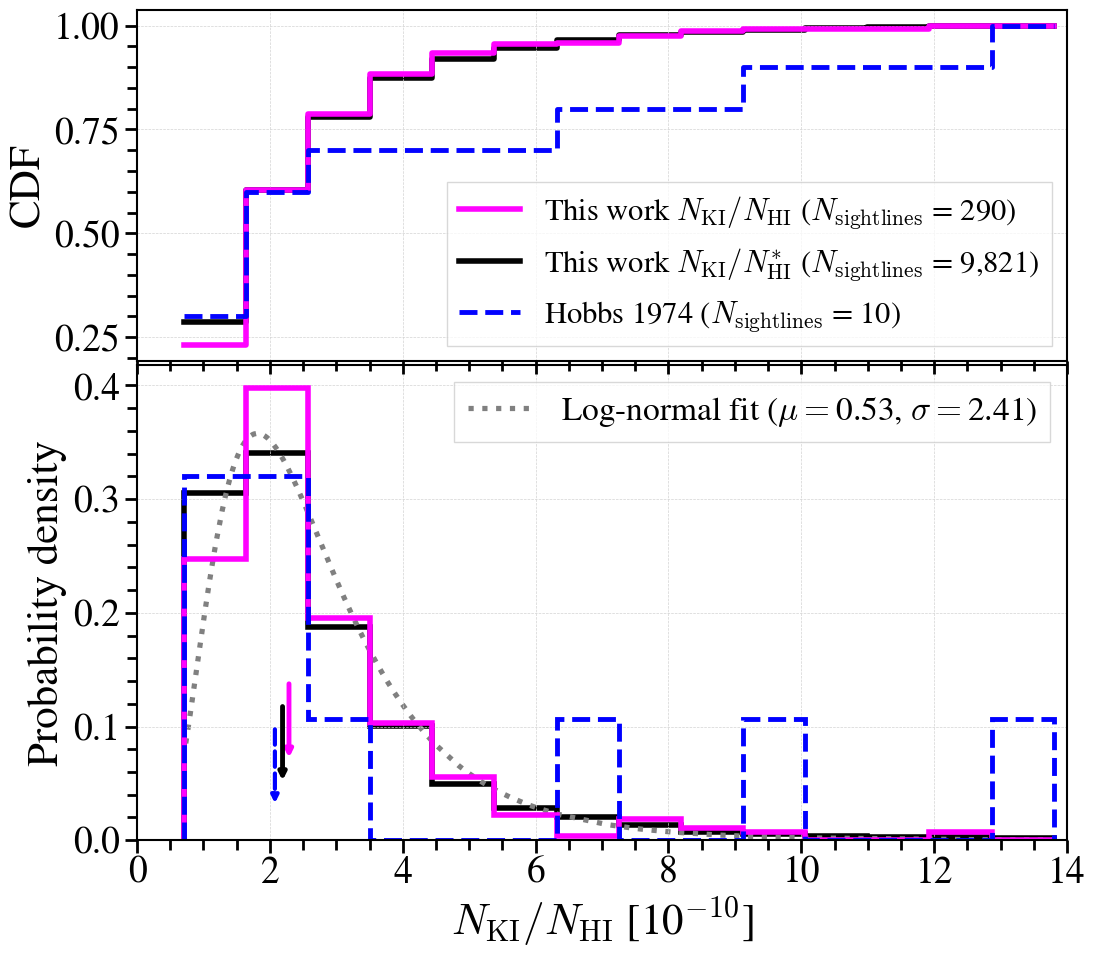}
\caption{Distributions of \ki/\hi\ abundance ratios (lower panel), i.e. \ki/\hi\ column density ratios, and their cumulative distribution functions (CDF, upper panel). The solid magenta line represents our current work (\nmatching\ lines of sight with optically-thick \hi\ column densities), whereas the dashed line shows data from \citealt[][]{Hobbs1974} (ten lines of sight). Arrows indicate the median values for each sample, at $\sim$2.3$\times$10$^{-10}$ and $\sim$2.1$\times$10$^{-10}$, respectively. The dotted curve represents the best-fit log-normal distribution ($\mu = 0.53$ and $\sigma = 2.41$) to the two combined \ki/\hi\ abundance ratios. For comparison, the solid black lines show the distributions for \nstars\ lines of sight with optically-thin \hi\ column densities, with its arrow indicating the corresponding median value $\sim$2.2$\times$10$^{-10}$.}
\label{fig:ki_hi_abundances}
\end{figure}

The distribution of the resulting \ki/\hi\ abundance ratios across our sample of \nmatching\ matching lines of sight is presented in Figure \ref{fig:ki_hi_abundances}, alongside a comparison to results derived from \nstars\ sightlines using optically-thin \hi\ column densities and to the findings of \citealt[][]{Hobbs1974} (ten lines of sight). For the \nmatching\ matched sightlines, our analysis reveals a \ki/\hi\ abundance ratio ranging from $0.8~\times~10^{-10}$ to $12.7~\times~10^{-10}$, with a median (mean, standard deviation) of $2.3 \times 10^{-10}$ ($2.8 \times 10^{-10}$, $1.8 \times 10^{-10}$). Additionally, by leveraging the CNM properties derived from the GASKAP-\hi\ absorption measurements \citep{Nguyen2024}, we estimate the ratio \NKI/\(N_{\mathrm{HI,\,CNM}}\) within our survey region, finding a median (mean and standard deviation) of 8.5 $\times$ 10$^{-10}$ (13.8 $\times$ 10$^{-10}$ and 16.0 $\times$ 10$^{-10}$).

Across the \nstars\ sightlines based on optically-thin assumption, the ratios \NKI/\NHIthin\ range from 
\(0.3 \times 10^{-10}\) to \(41.1 \times 10^{-10}\), with a median (mean, standard deviation) 
of \(2.2 \times 10^{-10}\) (\(2.8 \times 10^{-10}\), \(2.1 \times 10^{-10}\)). For the ten lines of sight examined by \citet[][see their Table 1]{Hobbs1974}, the reported \ki/\hi\ abundance ratios span from $0.7 \times 10^{-10}$ to $13.8 \times 10^{-10}$, with a median (mean, standard deviation) of $2.1 \times 10^{-10}$ ($4.2 \times 10^{-10}$, $4.2 \times 10^{-10}$). The excellent consistency between the statistical properties of the \ki/\hi\ abundance ratios derived from our significantly larger samples and those reported in literature provides strong support for the reliability of our \ki\ abundance estimates and suggests a degree of uniformity in these ratios across different lines of sight probed by the two studies.

Given a small sample size of ten sightlines in \cite{Hobbs1974}'s sample, we perform a Mann–Whitney U test and two-sample Kolmogorov-Smirnov (KS) test to assess whether the \cite{Hobbs1974} sample and our 290-sightline sample originate from the same underlying distribution. The Mann–Whitney U test, which determines if there is a statistically significant difference between the medians of two independent groups, produces a U statistic of 1415, $p-$value $=$ 0.9. The KS test produces a KS statistic of 0.3, $p-$value $=$ 0.5. Both tests indicate there are no statistically significant differences between the two \ki/\hi\ abundance ratio distributions. We then fit three candidate probability distributions -- exponential, log-normal, and gamma -- to the combined \ki/\hi\ abundance ratios (with opacity-corrected \NHI). The log-normal distribution with $\mu = 0.53$ and $\sigma = 2.41$ provides the best fit (KS statistic $=$ 0.06, $p-$value $=$ 0.18), while the exponential and gamma distributions are strongly rejected, with KS statistic $=$ 0.28 and 0.11, $p-$value $= 2 \times 10^{-21}$ and $2 \times 10^{-3}$, respectively.

\section{Conclusions and Future work}
    \label{conclusions}
    \label{sec:conclusions}
We have conducted a joint multi-wavelength analysis of the local neutral ISM with the use of two large-scale surveys: GASKAP for \hi\ absorption at 21 cm and GALAH optical spectroscopy for \ki\ absorption at 7699 \r{A}. Leveraging the Gaussian decompositions performed by \cite{Nguyen2024} for 462 GASKAP lines of sight with \hi\ absorption detections, together with GALAH measurements of interstellar \ki\ absorption in the spectra of nearby stars by \citet{Buder2024}, we investigated the kinematic relationship between CNM and neutral potassium in the Solar neighborhood at high Galactic latitudes ($-45^{\circ}, -25^{\circ}$) in the foreground of the Magellanic Clouds, as seen in Figure \ref{fig:all_src_locations}.

Our key findings are as follows:
\begin{enumerate}
\item Along 462 GASKAP lines of sight with \hi\ absorption detections, we identified \nmatching\ out of 462 ($\sim$63\%) overlapping lines of sight showing absorption in both \hi\ and \ki\ neutral species.

\item We find a strong kinematic correlation between these two tracers of cold neutral ISM in the Magellanic Cloud foreground. The small velocity offsets ($\sim 1$ \kms) observed between corresponding \hi\ and \ki\ absorption features indicate a close physical association, suggesting that \ki\ absorption likely arises within regions of cold hydrogen gas. In addition, a moderate positive correlation is found when examining \hi\ and \ki\ line-of-sight quantities, such as \ki\ column density compared to \hi\ column density, integrated optical depth, line-of-sight CNM column density or \hi\ brightness.

\item Our derived \ki/\hi\ abundance ratio is in excellent agreement with previous measurements reported by \citet{Hobbs1974}, with median ratios for the two samples, at $\sim$2.3$\times$10$^{-10}$ and $\sim$2.1$\times$10$^{-10}$, respectively.

\item Our GASKAP-GALAH synergy highlights the power of combining large-scale radio and optical surveys to gain new insights into the structure and composition of the local neutral ISM.
\end{enumerate}

Building upon strong \hi-\ki\ kinematic correlation found in this study, our future work will characterise the physical properties and spatial distribution of the local absorbing species across the sky, using \hi\ absorption detected in all existing Galactic absorption surveys. A key next step will be to estimate distances to the detected \hi\ gas features using the high-resolution GASKAP data in conjunction with detailed 3D dust maps of the Galactic ISM. Furthermore, we will establish a 3D field of H$\alpha$ emission by modelling the transport of H$\alpha$ photons and applying Gaussian inference to the well-constrained distances of nearby Solar neighborhood stars. By correlating the 3D distribution of \hi\ and \ki\ absorption with the derived 3D H$\alpha$ emission field, we will be able to investigate the ionisation state of hydrogen gas and explore the interplay between neutral and ionised components of the local ISM on larger scales.

\section*{Acknowledgements}
This scientific work uses data obtained from Inyarrimanha Ilgari Bundara / the Murchison Radio-astronomy Observatory. We acknowledge the Wajarri Yamaji People as the Traditional Owners and native title holders of the Observatory site. CSIRO’s ASKAP radio telescope is part of the Australia Telescope National Facility \hyperlink{https://ror.org/05qajvd42}{(https://ror.org/05qajvd42)}. Operation of ASKAP is funded by the Australian Government with support from the National Collaborative Research Infrastructure Strategy. ASKAP uses the resources of the Pawsey Supercomputing Research Centre. Establishment of ASKAP, Inyarrimanha Ilgari Bundara, the CSIRO Murchison Radio-astronomy Observatory and the Pawsey Supercomputing Research Centre are initiatives of the Australian Government, with support from the Government of Western Australia and the Science and Industry Endowment Fund.

This research was partially funded by the Australian Government through an Australian Research Council Australian Laureate Fellowship (project number FL210100039) to NMc-G and a Discovery Early Career Researcher Award (DE240100150) to SB. SS acknowledges the support provided by the University of Wisconsin-Madison Office of the Vice Chancellor for Research and Graduate Education with funding from the Wisconsin Alumni Research Foundation, and the NSF Award AST-2108370. JDS acknowledges funding by the European Research Council via the ERC Synergy Grant ``ECOGAL -- Understanding our Galactic ecosystem: From the disk of the Milky Way to the formation sites of stars and planets'' (project ID 855130).

This work made use of the Fourth Data Release of the GALAH Survey \citep{Buder2024}. The GALAH Survey is based on data acquired through the Australian Astronomical Observatory, under programs: A/2013B/13 (The GALAH pilot survey); A/2014A/25, A/2015A/19, A2017A/18 (The GALAH survey phase 1); A2018A/18 (Open clusters with HERMES); A2019A/1 (Hierarchical star formation in Ori OB1); A2019A/15, A/2020B/23, R/2022B/5, R/2023A/4, R2023B/5 (The GALAH survey phase 2); A/2015B/19, A/2016A/22, A/2016B/10, A/2017B/16, A/2018B/15 (The HERMES-TESS program); A/2015A/3, A/2015B/1, A/2015B/19, A/2016A/22, A/2016B/12, A/2017A/14, A/2020B/14 (The HERMES K2-follow-up program); R/2022B/02 and A/2023A/09 (Combining asteroseismology and spectroscopy in K2); A/2023A/8 (Resolving the chemical fingerprints of Milky Way mergers); and A/2023B/4 (s-process variations in southern globular clusters). We acknowledge the traditional owners of the Country on which the AAT stands, the Gamilaraay people, and pay our respects to Elders past and present. This paper includes data that has been provided by AAO Data Central (datacentral.org.au).

This work has made use of data from the European Space Agency (ESA) mission Gaia (https://www.cosmos.esa.int/gaia), processed by the Gaia Data Processing and Analysis Consortium (DPAC, https://www.cosmos.esa.int/web/gaia/dpac/consortium). Funding for the DPAC has been provided by national institutions, in particular the institutions participating in the Gaia Multilateral Agreement.

Finally, we thank the anonymous referee for the comments and suggestions that allowed us to improve the quality of our manuscript.

\textit{Software}: Astropy \citep{Pytorch2019}, Matplotlib \citep{MatplotlibHunter2007}, NumPy \citep{vanderWalt2011}, SciPy \citep{Virtanen2020}, Pandas \citep{mckinney2010data}.

\section*{Data Availability}
This paper includes archived data obtained through the CSIRO ASKAP Science Data Archive, CASDA \href{https://research.csiro.au/casda}{(https://research.csiro.au/casda)}.

All data related to the GALAH survey (https://www.galah-survey.org/) are publicly available at \url{https://cloud.datacentral.org.au/teamdata/GALAH/public/GALAH_DR4/}.

The GASKAP emission and absorption data used in this study, along with the fitted results and their associated uncertainties, are derived by \cite{Nguyen2024} and available at \href{https://github.com/GASKAP/HI-Absorption/tree/master/MW\_absorption}{GASKAP Github repository}.

All interstellar spectral data, their associated uncertainties, and the analysis notebooks used in this study are publicly available at \href{https://github.com/nv-hiep/gaga}{GASKAP GitHub repository} and \href{https://doi.org/10.5281/zenodo.17211553}{DOI 10.5281/17211553}.



\bibliographystyle{mnras}
\bibliography{references} 




\appendix

\section{Linking \hi\ gas to physical distance using 3D dust extinction and HI4PI}
\label{app_distances}
    In order to estimate physical distances to the local \hi\ gas, we utilise the recently released 3D map of interstellar dust extinction by \cite{Edenhofer2024} in conjunction with the HI4PI data cube \citep{hi4pi2016}.

The 3D dust map provides the distribution of differential extinction between 69 out to 1250 pc from the Sun. This map offers an angular resolution of $\sim$13.7 arcmin (HEALPix\footnote{The Hierarchical Equal Area isoLatitude Pixelisation \citep{Gorski2005}.} resolution parameter $N_\mathrm{side}$ = 256) and a distance resolution of 2 pc. Built upon Gaia mission data \citep{Gaia2023}, the 3D dust reconstruction employs a Gaussian process to model differential dust extinction in spherical coordinates for 54 million nearby stars, assuming a spatially smooth distribution of dust extinction. The differential extinction is given in unitless values, defined by \cite{Zhang2023} (hereafter ZGR23) as ($dA_\mathrm{ZGR23} / 1$ pc). These extinction units can be converted into extinction at a specific wavelength using the publicly available ZGR23 extinction curve\footnote{ZGR23 extinction curve: \href{https://doi.org/10.5281/zenodo.7692680}{DOI 10.5281/zenodo.7811871}}.

For \hi\ emission, we employ the HI4PI data cube, constructed from the Effelsberg-Bonn \hi\ Survey (EBHIS; \citealt{Winkel2010,Kerp2011}) and the Galactic All-Sky Survey (GASS; \citealt{McClureGriffiths2009}) single-dish observations. The combined set has an angular resolution of 16.2 arcmin and a sensitivity of 43 mK per 1.29 \kms\ spectral resolution. This angular resolution is close to the 14 arcmin resolution of the 3D dust map. Assuming \hi\ kinematics as a proxy for the gas physical distances, these comparable resolutions potentially allow for an informative morphological comparison between dust extinction and \hi\ emission structures.

\begin{figure}
\includegraphics[width=1.0\linewidth]{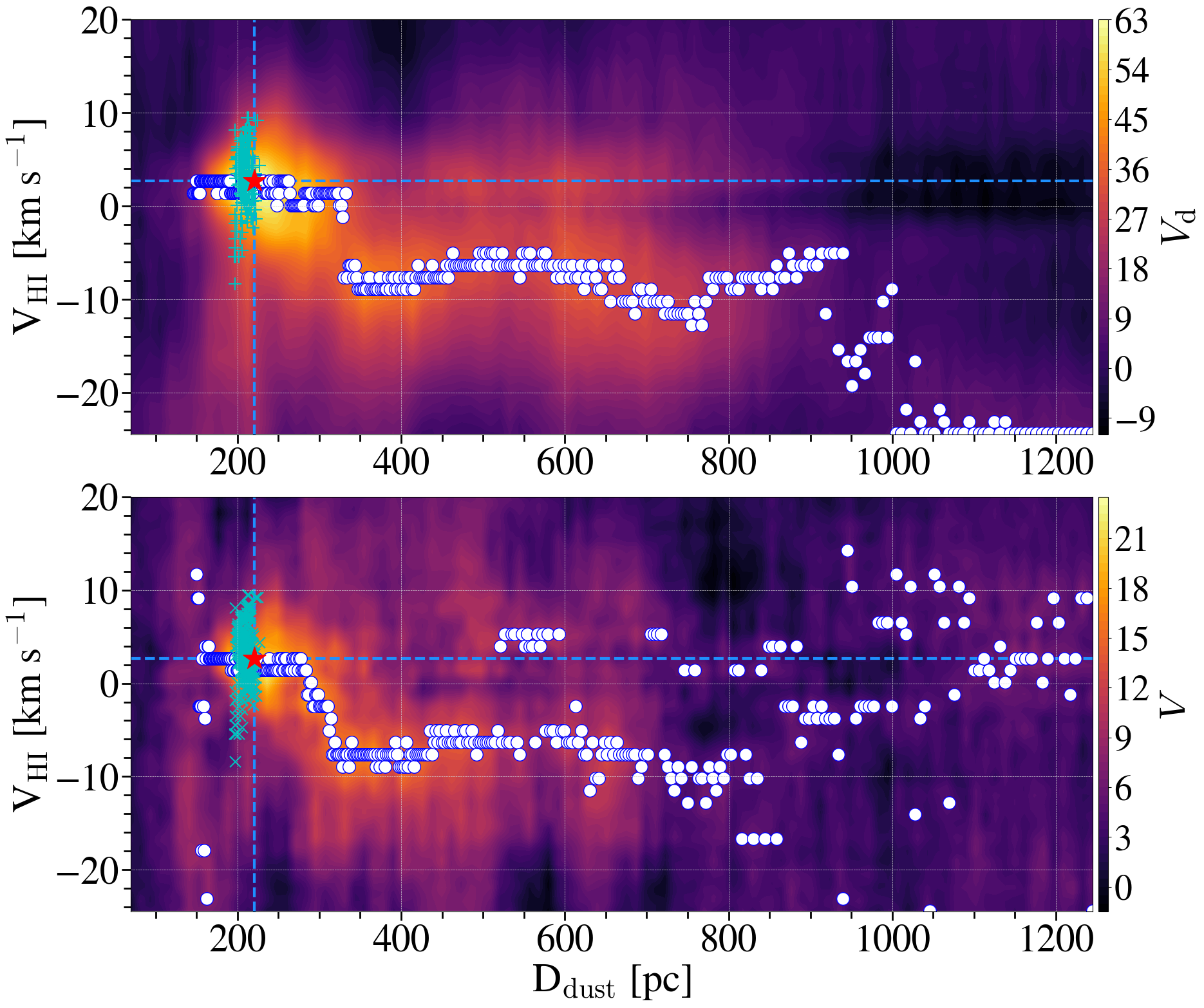}
\caption{Morphological correlation
between 3D dust distance channels and \hi\ velocity channels for GASKAP's Galactic region $l$ = (258, 316)$^{\circ}$, $b$ = ($-$50, $-15$)$^{\circ}$. The correlation metrics are the direction-sensitive projected Rayleigh statistic ($V_\mathrm{d}$ in upper panel) and the projected Rayleigh statistic ($V$ in lower panel). In both panels, circle markers indicate the velocity with the highest Rayleigh statistic values for each distance channel, star markers at the intersection of the dashed lines (at $D_\mathrm{dust} =$ 220.3 pc, $V_\mathrm{HI} =$ 2.7 \kms) indicates the velocity-distance pairs with the highest Rayleigh statistic across all velocity and distance ranges. Plus markers (``$+$'') show velocity-distance matching from a simple matching between velocity at peak \hi\ optical depth $\tau_\mathrm{peak}$ and distance at peak dust extinction along 462 GASKAP lines of sight with \hi\ absorption detections. Cross markers (``$\times$'') show velocity-distance matching from a simple matching between velocity at peak \hi\ brightness temperature $T_\mathrm{b,peak}$ and distance at peak dust extinction along the same \hi\ absorption detection lines of sight.}
\label{fig:hog_V_planes}
\end{figure}

We then apply the Histogram of Oriented Gradients (HOG) method \citep{Soler2019} to the gas and dust cubes to identify spatial correlations between the two ISM tracers. This method is originally designed to characterise the similarities in the emission distribution based on gradient orientations. In this work, HOG analysis quantifies the morphological similarity between the 2D plane-of-the-sky distribution of the 3D dust extinction and \hi\ emission. This is done by comparing distance channels ($D_\mathrm{dust}$ in pc) from dust map with velocity channels ($V_\mathrm{HI}$ in \kms) from \hi\ data. A velocity channel map and a distance channel map are considered morphologically similar if their gradients are mainly parallel, and dissimilar if they exhibit randomly oriented. Specifically, for each pair of \hi\ velocity and dust distance maps, the HOG method produces the projected Rayleigh statistic ($V$, see \citealt{Soler2019}) and the direction-sensitive projected Rayleigh statistic ($V_\mathrm{d}$, see \citealt{Soler2025}) to quantify the relationship between the 3D dust and \hi\ emission.

The Rayleigh statistic ($V$ or $V_\mathrm{d}$) assesses non-uniformity in a distribution of angles around a specific direction. A positive Rayleigh statistic indicates that the gradients are primarily parallel, signifying a morphological similarity between the \hi\ emission and dust extinction in the corresponding velocity-distance channel pair. In contrast, a negative Rayleigh statistic suggests that the gradients are mostly antiparallel, leading to an anti-correlation between the dust extinction and \hi\ emission intensity.

\begin{figure}
\includegraphics[width=1.0\linewidth]{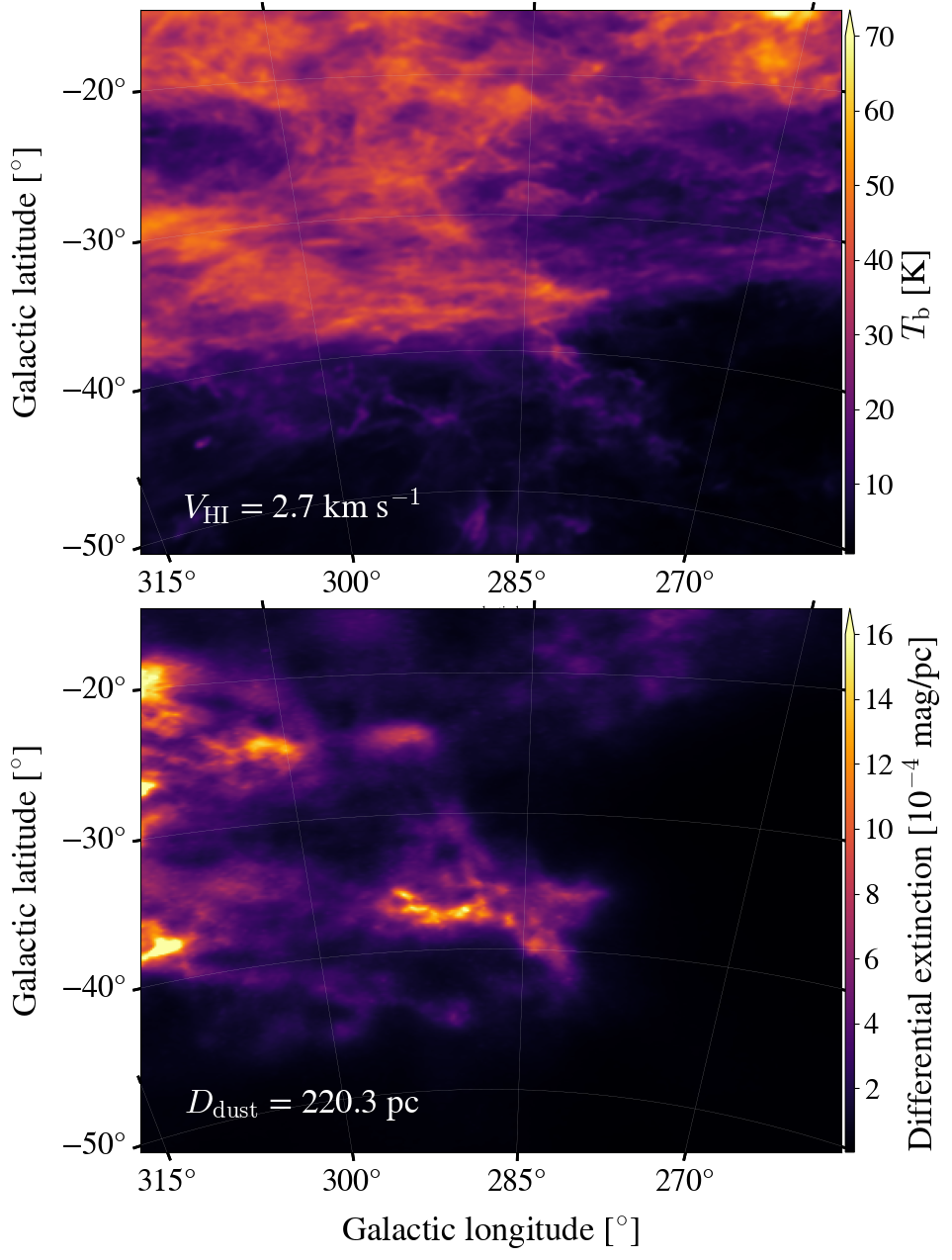}
\caption{A pair of \hi\ velocity and dust distance channel maps with the strongest morphological correlation, as determined by the HOG method. These two maps correspond to the highest $V_\mathrm{d}$ (and $V$) in the comparison between the \hi\ emission and the 3D dust extinction presented in Figure \ref{fig:hog_V_planes}.
{\it Top panel}: HI4PI emission at velocity $V_\mathrm{HI}$ $\approx$ 2.7 \kms. {\it Bottom panel}: Dust differential extinction at distance $D_\mathrm{dust}$ $\approx$ 220.3 pc obtained from \citet{Edenhofer2024}.
}
\label{fig:hi_dust_max_v}
\end{figure}

In this Section, we consider the local gas in the high galactic-latitude region $l =$ (258, 316)$^{\circ}$, $b =$ ($-$50, $-15$)$^{\circ}$ with velocity range from $-$25 to $+$20 \kms, and the dust distance from 69 to 1250 pc from the Sun. Figure \ref{fig:hog_V_planes} illustrates the HOG morphological correlation between velocities from \hi\ emission observations and distances from 3D dust extinction reconstruction for our region of interest. The upper panel displays the velocity-distance correlations with the direction-sensitive projected Rayleigh statistic $V_\mathrm{d}$ colour-coded, the lower panel for the same correlations but for the projected Rayleigh statistic $V$. Following \cite{Soler2025}, $V_\mathrm{d}$ values at $\sim$0 indicates a random orientation and low morphological correlation between the gradients of the two tracers. Values of $V_\mathrm{d} > 2.87$ suggest mostly parallel gradients and a significant morphological correlation, whereas $0 < V_\mathrm{d} < 2.87$ indicate mostly antiparallel gradients. Within each distance channel, white dots mark the velocity with the highest Rayleigh statistic values. Star markers denote the velocity-distance pairs with the highest Rayleigh statistic across all velocity and distance ranges. In both panels, they are exactly at the same velocity-distance pairs $D_\mathrm{dust}$ = 220.3 pc, and $V_\mathrm{HI}$ = 2.7 \kms, although the white dots are more dispersed for the $V$ Rayleigh statistic values, in particular at closer distances $D_\mathrm{dust} < 170$ pc. Figure \ref{fig:hi_dust_max_v} presents an example of \hi\ emission along with dust extinction distribution, for the velocity-distance pair that obtains the highest morphological correlation as identified by the peak $V_\mathrm{d}$ and $V$ values in Figure \ref{fig:hog_V_planes}.

In addition to the HOG approach, we performed a simple velocity-distance matching. This involved comparing the velocity at peak \hi\ optical depth ($\tau_\mathrm{peak}$) with the distance at peak dust extinction along 462 GASKAP lines of sight with \hi\ absorption detections. We also matched the velocity at peak \hi\ brightness temperature ($T_\mathrm{b,peak}$) with the distance at peak dust extinction along the same lines of sight. These simple matchings are represented by plus (``$+$'') and cross (``$\times$'') markers in the upper and lower panels of Figure \ref{fig:hog_V_planes}, respectively. Both HOG analysis and simple velocity-distance matching indicate that the distances to local \hi\ gas in the direction of the Magellanic Cloud foreground range from 160 pc to 260 pc, with a likely distance $\sim$220 pc and a corresponding \hi\ gas velocity of $\sim$3 \kms.


\bsp	
\label{lastpage}
\end{document}